\documentclass[times,twocolumn]{aastex631}
\usepackage{amsmath}
\usepackage{amssymb}

\newcommand{\bracketfunc}[3]{\left(\frac{#1}{#2}\right)^{#3}}

\newcommand{\fargo}{{\sc \tt FARGO} }
\newcommand{\mpl}{M_{\rm p}}
\newcommand{\mstar}{M_{\ast}}
\newcommand{\rp}{R_{\rm p}}

\newcommand{\sigmagas}{\Sigma_{\rm g}}

\newcommand{\vradg}{\rm{v}_{R,{\rm g}}}
\newcommand{\vphig}{\rm{v}_{\phi,{\rm g}}}
\newcommand{\vradd}{\rm{v}_{R,{\rm d}}}

\newcommand{\h}{h}
\newcommand{\hp}{h_{\rm p}}
\newcommand{\cs}{c_s}

\newcommand{\mdisk}{M_{\rm disk}}
\newcommand{\omegak}{\Omega_{\rm K}}
\newcommand{\omegakp}{\Omega_{\rm K,p}}

\newcommand{\dsize}{s_d}
\newcommand{\dsizezero}{s_{d,0}}
\newcommand{\facc}{f_{\rm eff}}
\newcommand{\vk}{v_{\rm K}}
\newcommand{\st}{St \ }
\newcommand{\etamin}{\eta_{\rm min}}
\newcommand{\etaminovertime}{\overline{\eta}_{\rm min}}
\newcommand{\tstart}{t_{\rm b,leak}}
\newcommand{\tend}{t_{\rm b,end}}
\newcommand{\tbumpform}{t_{\rm b,form}}
\newcommand{\tbumpformmax}{t_{\rm b,max}}
\newcommand{\fmdust}{F_{\rm M,dust}}
\newcommand{\tautypeI}{\tau_{\rm type~I}}
\newcommand{\tauwgap}{\tau_{\rm type~II}}
\newcommand{\taudrift}{\tau_{\rm d,drift}}
\newcommand{\taugapform}{\tau_{\rm gap}}
\newcommand{\tdelay}{t_{\rm delay}}
\newcommand{\rring}{R_{\rm ring}}
\newcommand{\sigmacrit}{\Sigma_{\rm g,crit}}
\newcommand{\mdiskcrit}{M_{\rm disk,crit}}
\newcommand{\miso}{M_{\rm iso}}

\received{\today}
\revised{\today}
\accepted{\today}
\submitjournal{ApJ}

\shorttitle{Dust rings as a footprint of planet formation in a protoplanetary disk}
\shortauthors{K.D. Kanagawa et al.}

\begin{document}

\title{Dust rings as a footprint of planet formation in a protoplanetary disk}

\correspondingauthor{Kazuhiro D. Kanagawa}
\email{kazuhiro.kanagawa.d@vc.ibaraki.ac.jp}

\author[0000-0001-7235-2417]{Kazuhiro D. Kanagawa}
\affiliation{College of Science, Ibaraki University, 2-1-1 Bunkyo, Mito, Ibaraki 310-8512, Japan}

\author{Takayuki Muto}
\affiliation{Division of Liberal Arts, Kogakuin University, 1-24-2 Nishi-Shinjuku, Shinjuku-ku, Tokyo 163-8677, Japan}

\author{Hidekazu Tanaka}
\affiliation{Astronomical Institute, Tohoku University, Sendai, Miyagi 980-8578, Japan}



\begin{abstract}
Relatively large dust grains (referred to as pebbles) accumulate at the outer edge of the gap induced by a planet in a protoplanetary disk, and a ring structure with a high dust-to-gas ratio can be formed.
Such a ring has been thought to be located right outside of the planet orbit.
We examined the evolution of the dust ring formed by a migrating planet, by performing two-fluid (gas and dust) hydrodynamic simulations.
We found that the initial dust ring does not follow the migrating planet and remains at the initial location of the planet in the cases with a low viscosity of $\alpha \sim 10^{-4}$.
The initial ring is gradually deformed by viscous diffusion, and a new ring is formed in the vicinity of the migrating planet, which developes from the trap of the dust grains leaking from the initial ring.
During this phase, two rings co-exist outside the planet orbit.
This phase can continue over $\sim 1$~Myr for a planet migrating from 100~au.
After the initial ring disappears, only the later ring remains.
This change in the ring morphology can provide clues as to when and where the planet was formed, and is the footprint of the planet.
We also carried out simulations with a mass-growing planet. 
These simulations show more complex asymmetric structures in the dust rings.
The observed asymmetric structures in the protoplanetary disks may be related to a migrating and mass-growing planet. 
\end{abstract}


\section{Introduction} \label{sec:intro}
Recent observations by, for example, the Atacama Large Millimeter Array (ALMA) have revealed many protoplanetary disks with structures such as rings/gaps \citep[e.g.,][]{ALMA_HLTau2015,Tsukagoshi2016,Nomura_etal2016,Long2018,VanderMarel2019}, spiral arms \citep[e.g.,][]{Muto2012,Huang2018,Wolfer2021,Xie2021}, cavities \citep[e.g.,][]{Dong2018_MWC758,Kudo2018,Hashimoto2021}, and vortexes \citep[e.g.,][]{Fukagawa2013,Soon2019,Yamaguchi2020,Hashimoto_Dong_Muto2021}.
Although there are several possible mechanisms to form these substructures, for instance, for gap/ring structures, snow line \citep{Zhang_Blake_Bergin2015}, sintering effects \citep{Okuzumi_Momose_Sirono_Kobayashi_Tanaka2016,Hu2019,Hu2021}, secular gravitational instabilities \citep{Takahashi_Inutsuka2014,Takahashi_Inutsuka2016a}, and disk-wind \citep{Takahashi_Muto2018}, the disk--planet interaction is one of the most convincing mechanisms to be the origin of the observed substructure.
The planet gravitationally interacts with surrounding gas and exchanges the angular momentum through spiral waves \citep{Goldreich_Tremaine1979,Goldreich_Tremaine1980,Lin_Papaloizou1979}.
As a result of the strong interaction, a giant planet can form a density gap \citep{Lin_Papaloizou1986a,Kley1999,Nelson_Papaloizou_Masset_Kley2000,Crida_Morbidelli_Masset2006,Duffell_MacFadyen2013,Fung_Shi_Chiang2014,Kanagawa2015a}, and an even larger planet (or companion) can open a cavity \citep{Artymowicz_Lubow1994,Gunther_Kley2002,Kley_Haghighipour2014,Thun_Kley_Picogna2017,Miranda_Munoz_Lai2017}.
When the planet carves a density gap, relatively large dust grains (known as pebbles) are accumulated at the outer edge of the gap, and the dust ring is formed \citep{paardekooper2004,Muto_Inutsuka2009b,Zhu2012,Pinilla_Ovelar_Ataiee_Benisty_Birnstiel_Dishoeck_Min2015,Kanagawa_Muto_Okuzumi_Taki_Shibaike2018,Kanagawa2019}.
These substructures could provide clues as to where and when planets form in the protoplanetary disks.

Thanks to the ALMA, a variety of the ring structures on protoplanetary disks have been revealed;
for instance, the disks with the dust ring outside the large inner cavity, known as transition disks, such as Sz~91 \citep{Tsukagoshi_Sz91_2019}, and the smooth filled disks with multiple dust rings, such as HL~Tau.
In particular, the disks with two outer high-contrast rings stand out among the disks with multiple rings, e.g., AS~209, HD~143006, and HD~163296 \citep{Fedele_etal2017,Fedele2018,Huang_DSHARP}.
Most of the rings can be fitted by a simple gaussian annual shape, and such symmetric rings can be interpreted by the mechanisms described above, such as the disk-planet interaction and snow line scenarios.
However, some of the rings have asymmetric structures within the asymmetric ring, as can be seen in the disks of HD~143006, HD~34282, and HD~100453 \citep{vanderPlas2017, Huang_DSHARP,Francis_vanderMarel2020}.
More intriguingly, the disk of Elias~24 has one high-contrast ring and a flat intensity region outside the high-contrast ring \citep{Huang_DSHARP}.
These complex rings are difficult to interprete based on the simple planet-disk interaction and snow-line scenarios.
A more sophisticated model would be required to explain these complex rings.

Planetary migration is one of the factors that can make rings induced by the planet complex.
\cite{Dong_Li_Chiang_Li2017} reported that a planet form two or more rings in the cases with a low viscosity of $\alpha \lesssim 5\times 10^{-5}$, where $\alpha$ is the viscosity parameter \citep{Shakura_Sunyaev1973}, and found that the locations of the rings depend on the planetary migration.
\cite{Weber2019} also investigated the effect of the migration in the low viscosity disk with $\alpha=10^{-5}$ and showed the consistent result with the work of \cite{Dong_Li_Chiang_Li2017}.
\cite{Meru_Rosotti_Booth_Nazari_Clarke2018} investigated the dust ring induced by a migrating planet when $\alpha=10^{-3}$ and found that the dust ring can be formed on both sides of the planetary orbit, depending on the radial drift speed of dust grains.
In the previous studies, the dust rings were always formed in the vicinity of the planet and follow the migrating planet, and the studies for the low-viscosity disk did not consider a long-term evolution of the rings with a significant planetary migration.
However, in a disk with a low viscosity ($\alpha \lesssim 10^{-4}$),  the local viscous timescale, $h^2/\nu$, can be much longer than the migration timescale of the planet, where $h$ is the scale height and $\nu$ is the kinetic viscosity.
In such a disk, the gas structure perturbed by the planet may not be restored to the unperturbed one even after the planet moves far away, which could form more complex rings.

In this study, we investigate the relation between the locations of the ring and the planet, by carrying out two-fluid (gas and dust) two-dimensional hydrodynamic simulations in disks with relatively low viscosity, $\alpha \sim 10^{-4}$.
Our simulations showed that the morphology of the dust rings varies as the planet migrates, and two rings can co-exist outside the orbit of the migrating planet, in such a low viscosity disk.
We also found more complex asymmetric structures in dust rings for cases of a mass-growing planet.
In Section~\ref{sec:model}, we describe our setup of hydrodynamic simulations, and we present results of the cases with a fixed mass planet and mass-growing planet in Section~\ref{sec:results}.
In Section~\ref{sec:discussion}, we discuss the relation between the locations of the ring and planet and implications for the formation of the debris disks and planets.
In Section~\ref{sec:summary}, we summarize our results.

\section{Model and numerical method} \label{sec:model}
We simulate a radial migration of a planet in a protoplanetary disk by using two-fluid (gas and dust) hydrodynamic simulations.
The simulations are done using the open source hydrodynamic code \fargo \citep{Masset2000} with the implementation of dust components.
The basic equations and treatment of the dust components are described in the previous papers \citep{Kanagawa_Ueda_Muto_Okuzumi2017,Kanagawa_Muto_Okuzumi_Taki_Shibaike2018,Kanagawa2019}.
Here, we briefly summarized the setup of the simulations.

We solved continuous equations and equations of motions for gas and dust components in two-dimensional coordinates ($R,\phi$) by assuming a geometrically thin and non-self-gravity disk.
We also assume a simple, locally isothermal equation of state.
The central star is located at the origin of the coordinate, and the mass of the central star $\mstar$ is constant.
We take into account the feedback of the dust grains on the gas, and, hence, the gas is affected by the dust in a similar manner as the dust drift \citep{Kanagawa_Ueda_Muto_Okuzumi2017,Kanagawa_Muto_Okuzumi_Taki_Shibaike2018}.

We examine the evolution of the dust distribution perturbed by a single migrating planet initially located at an orbital radius of $\sim 100$~au.
We include planet migration relevant to a torque exerted from the surrounding gas and dust grains in the disk \citep{Kanagawa2019}. 
In our simulations, the initial orbital radius, $R_0$, and the mass of the central star, $\mstar$, are the units of length and mass, respectively. 
The surface density is normalized by $\mstar/R_0^2$.
We use $t_0 = 2\pi/\omegak(R_0)$ as the unit of time, where $\omegak$ is the Keplerian angular velocity.
Note that when $R_0 = 100$~au and $\mstar=1M_{\odot}$, $t_0$ is 1000~yr. 

The computational domain is $0.3R_0$ to $2.4R_0$.
The radial domain is divided by 2048 meshes with logarithmic spacing, and the azimuthal domain is divided by 2048 meshes with equal spacing. 
We assume the disk with a constant disk aspect ratio $h/R$ over the computational domain (which we denote $H_0$ in the following), where $h$ is a disk scale height.
We adopt the $\alpha$-prescription of \cite{Shakura_Sunyaev1973} for the kinetic viscosity $\nu$.
Hence, the kinetic viscosity is given by $\nu=\alpha h^2\omegak$, and we assume a constant value of $\alpha$ over the computational domain.
Considering the steady state of the viscous accretion disk \citep[c.f.][]{Pringle1981}, we adopted the power-law distribution as the initial distribution of the gas surface density, namely, $\sigmagas = \Sigma_0 (R/R_0)^{-1/2}$.
In the following, the subscript 'g' indicates the value of the gas, and similarly, the subscript 'd' indicates the value of the dust component.
The radial velocity of the gas is given by $\vradg = -3\nu/R$.
The azimuthal velocity of the gas is given by $\vphig = R\omegak\sqrt{1-2\eta}$, where $\eta=-(1/2)(h/R)^2 d\ln (\cs^2\sigmagas) /d\ln R$ and $\cs$ is the isothermal sound speed.
For the dust grains, an initial gas-to-dust ratio is fixed be to 0.01, and the radial and azimuthal velocities are given by zero and $R\omegak$, respectively.
The fiducial value of $\Sigma_0$ is $10^{-3}$, and it corresponds to $0.9 \mbox{ g/cm}^2$ when $\mstar=1M_{\odot}$ and $R_0=100 \mbox{ au}$.
We adopt the same inner and outer boundary as those adopted in \cite{Kanagawa2019}.
Around the inner boundary, we set wave-killing zones with the width of 0.1 $R_0$.
For the planetary gravitational potential, we set the smoothing parameter to $0.6\hp$, where $\hp$ is the disk scale height at the planet orbital radius $\rp$.
In the following, the subscript $p$ indicates the value at $\rp$.

In this study, because we focus on the evolution of relatively large dust grains  (pebbles), we do not consider any dust growth and fragmentation processes.
For simplicity, we just assume a constant size of the dust grains.
The fiducial size of the dust grains $\dsize$ is $0.1\dsizezero$, where $\dsizezero$ is defined by
\begin{align}
\dsizezero=5.8 \bracketfunc{\Sigma_0}{0.9 \mbox{ g/cm}^2}{} \bracketfunc{\rho_d}{1 \mbox{ g/cm}^3}{-1} \mbox{ cm},
\label{eq:dsizezero}
\end{align}
and $\rho_d$ is the internal density of the dust grains.
Hence, the size of $0.1\dsizezero$ indicates millimeter-sized dust grains when $\Sigma_0=0.9 \mbox{ g/cm}^2$.
Adopting the Epstein regime, we can express the Stokes number of the dust grains as
\begin{align}
St &= \frac{\pi \dsize \rho_d}{2\sigmagas}.
\label{eq:st}
\end{align}
Note that $\dsizezero$ corresponds to the size that the Stokes number is equal to unity at $R_0$ at the initial distribution.

In the modern core-accretion model, a planet can develop by capturing pebbles, referred to as pebble accretion.
The mass growth timescale due to the pebble accretion is proportional to $\mpl^{1/3}$ \citep{Lambrechts_Johansen_Morbidelli2014}, whereas the timescale of Type~I migration is proportional to $\mpl^{-1}$ \citep{Tanaka_Takeuchi_Ward2002}.
Thus, for planets heavier than a certain mass, the migration timescale is much shorter than the growth time \citep{Johansen_Ida_Brasser2019}.
In this case, the fixed planet mass during the migration could be a good approximation.
When its mass reaches that is known as the pebble-isolation mass, the planet forms a pressure bump outside the planetary orbit, and and the pebble accretion is halted due to the trapping at the pressure bump, which is called pebble-isolation \citep{Morbidelli_Nesvorny2012,Lambrechts_Johansen_Morbidelli2014}.
After the pebble-isolation, the planet can capture an amount of gas from the protoplanetary disk, as its atmosphere shrinks, known as runaway gas accretion.
The duration between the pebble-isolation and the onset of the runaway gas accretion could depend on the opacity of the atmosphere and accretion history of the planetary core \citep{Pollack_etal1996,Ikoma_Nakazawa_Emori2000,Hubickyj_Bodenheimer_Lissauer2005}, but it is not fully understood.
Hence, we consider two extreme cases, the cases with and without growth of the planet mass.
In Section~\ref{subsec:ring_wacc}, we consider the gas accretion of the planet, while we consider the planet with a fixed mass in the rest of the paper.

Gas accretion has been investigated by many previous works \citep[e.g.,][]{Kley1999,Bate_Lubow_Ogilvie_Miller2003,D'Angelo_Kley_Henning2003,Bodenheimer_DAngelo_Lissauer_Fortney_Saumon2013,Szulagyi_Morbidelli_Crida_Masset2014,Tanigawa_Tanaka2016,Durmann_Kley2017,Tanaka_Murase_Tanigawa2019,Li_Chen_Lin_Zhang2021}.
In this study, however, instead of adopting the detailed model of the accretion process, we investigate the effects of gas accretion on the evolution of the dust ring by using the simple model developed by \cite{Kley1999} with the following accretion efficiency, which is implemented in \fargo.
At each time step, the gas density in cells within the Hill radius $R_H=\rp (\mpl/3\mstar)^{1/3}$ around the planet is reduced by a fraction of $\facc f_{\rm acc} \Delta t \omegakp$, where $\Delta t$ is the time step, and $\facc$ and $f_{\rm acc}$ are the free parameters.
The free parameter $f_{\rm acc}$ defines the size of the accretion region (for a detailed form of $f_{\rm acc}$, see \cite{Kley1999}), and $\facc$ determines the efficiency of the gas accretion.
We adopt three values of $\facc$ as described in Section~\ref{subsec:ring_wacc}.
Moreover, when the mass of the planet is small, the accretion rate calculated by the above method can be larger than the accretion rate determined by the Kelvin-Helmholtz quasi-static contraction \citep{Pollack_etal1996,Ikoma_Nakazawa_Emori2000} which is given by
\begin{align}
\dot{M}_{\rm p,KH} &\simeq \frac{\mpl}{\tau_{\rm KH}},
\label{eq:mdot_kh}
\end{align}
where $\tau_{\rm KH}$ is the contraction timescale.
It can be written as \citep{Ikoma_Nakazawa_Emori2000}
\begin{align}
\tau_{\rm KH} &\simeq 10^{2} \bracketfunc{\mpl}{M_{\oplus}}{k_1}\bracketfunc{\kappa_{g}}{1\mbox{cm}^2 \mbox{g}^{-1}}{k_2} \mbox{Myr},
\label{eq:khtime}
\end{align}
with $k_1\simeq -2.5$ to $-3.5$ and $k_2\simeq -1$.
For simplicity, we adopt $k_1=-3$, $k_2=-1$, and $\kappa_{\rm g}=1 \mbox{cm}^2/\mbox{g}$, and moreover, we adopt $1 M_{\odot}$ as the unit of the mass in Equation~(\ref{eq:khtime}).
If the accretion rate is larger than that given by Equation~(\ref{eq:mdot_kh}), we adjust $\facc$ to let the accretion rate be that given by Equation~(\ref{eq:mdot_kh}).

Aside from the above gas accretion, to avoid an abrupt entry of the planet at the beginning of the simulations, we gradually increase the mass of the planet to the given initial mass, by the function of $\mpl \sin \left[\pi t^2/(2 \tdelay^2)\right]$ for $t<\tdelay$, and the planet mass reaches $\mpl$ at $t=\tdelay$, where $\tdelay$ is a free parameter to determine the duration of the initial mass growth.
This delay time may be regarded as the timescale of the core growth due to the pebble accretion.
We adopt $\tdelay = 10 \ t_0$ as the fiducial value, which is much shorter than the timescales of the planetary migration and the formation of the pressure bump.
In Appendix~\ref{sec:effect_numerical_setup}, carrying out the simulations with longer $\tdelay$, namely $\tdelay \sim 100 \ t_0$, we confirmed that our results are not affected by the choice of $\tdelay$.

\section{Results} \label{sec:results}
\subsection{Formation and destruction of dust rings} \label{subsec:formation_and_distruction_dustring}
\begin{figure*}
	\begin{center}
		\resizebox{0.98\textwidth}{!}{\includegraphics{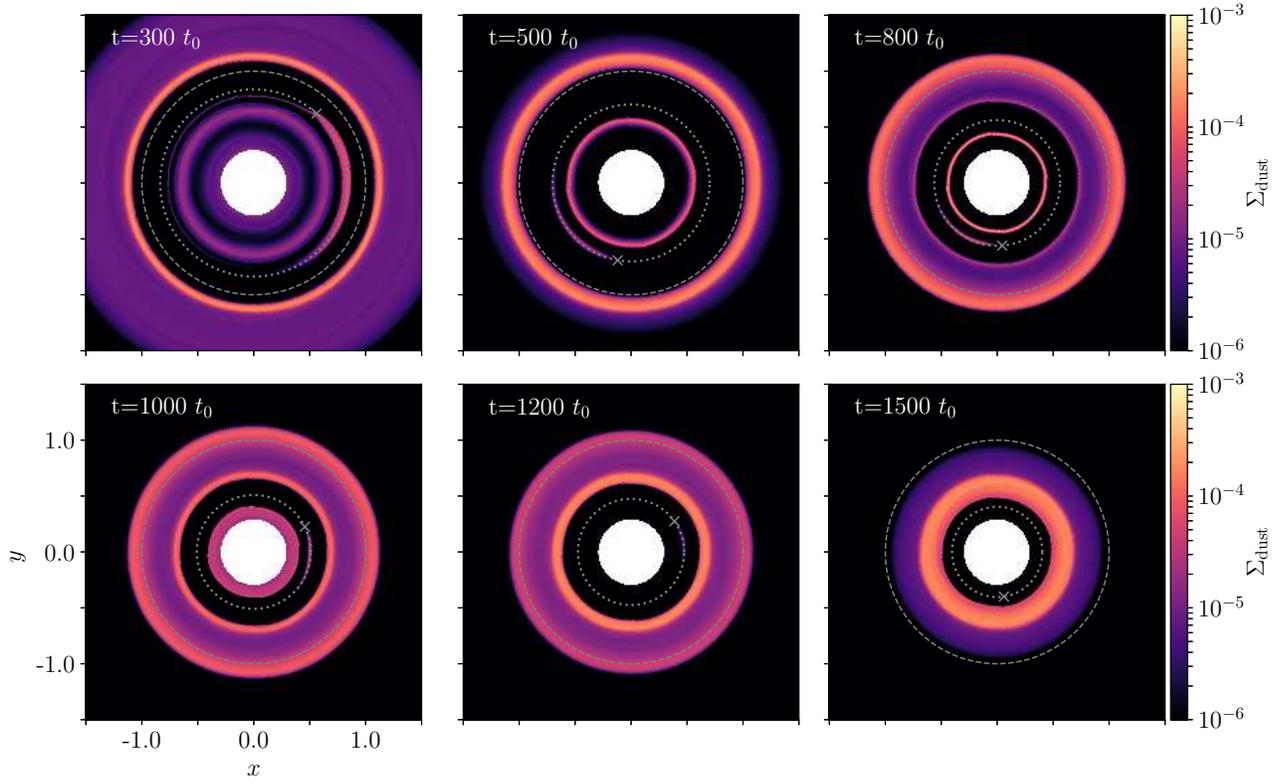}}
		\caption{
 		Time variation of dust surface density distribution in the case of $\mpl/\mstar=10^{-4}$, $H_0=0.05$, and $\alpha=3\times 10^{-4}$.
 		The outer dashed circle indicates the initial orbital radius of the planet at $R_0$, and the inner dotted circle denotes the orbital radius of the planet at the particular time of each panel.
 		In this case, $\tbumpform=60 \ t_0$, $\tbumpformmax=210 \ t_0$, $\tstart=570 \ t_0$, and $\tend=1500 \ t_0$ (for these characteristic times, see Section~\ref{subsec:scaling}).
 		The cross in each panel indicates the location of the planet.
		\label{fig:dens2d_q1e-4_h5e-2_a3e-4}
		}
	\end{center}
\end{figure*}
We first show the results with the fixed mass of the planet during the simulation, and in Section~\ref{subsec:ring_wacc} we show the results with a mass-growing planet considering the gas accretion.
Figure~\ref{fig:dens2d_q1e-4_h5e-2_a3e-4} shows a time series of two-dimensional distributions of dust grains in the case of $\mpl/\mstar=10^{-4}$, $H_0=0.05$, and $\alpha=3\times 10^{-4}$.
In the figure, we denote the initial orbit of the planet ($R=R_0$) as a dashed circle, and the present orbit of the planet ($R=\rp$) as a dotted line.
Note that the dotted circle is always located inside the dashed circle because of the inward migration of the planet.
In the early phase, namely, $t<300\ t_0$, the initial dust ring forms at the outer edge of the gap induced by the planet.
Though the planet migrates inward due to disk--planet interaction, the initial ring does not move, but becomes gradually wider as time passes, as can be seen from the series of the plots of $t=300 \ t_0$, $500 \ t_0$, and $800 \ t_0$.
Moreover, from $t=500 \ t_0$ to $800 \ t_0$, the bright and narrow secondary ring is formed at the inside of the planetary orbit which is associated with the secondary gap of the planet \citep{Dong_Li_Chiang_Li2017,Bae_Zhu_Hartmann2017,Dong_Li_Chiang_Li2018}.
This secondary ring starts to be deformed around $t=800 \ t_0$ and it vanishes at $t=1200\ t_0$, but it is due to the effect of the inner boundary and it can survive at $t=1200 \ t_0$ if the smaller radius of the inner boundary is adopted (see Appendix~\ref{sec:effect_numerical_setup}).
Therefore, in this study, we focus on the rings located at the outside of the planet orbit, which could be associated with the initial location and present position of the planet.
After $t=800 \ t_0$, the outer initial ring gradually deforms and a new dust ring (the later ring) is formed close to the planetary location.
The later ring becomes thicker as the outer initial ring vanishes ($t=1200 \ t_0$), and, finally, only the later ring remains ($t=1500 \ t_0$).
It is worth noting that though the deep gap of the dust grains is formed quickly, at the coorbital radius of the planet, dust grains remain for a longer time and become thinner with time and finally vanish at $t \sim 1000 \ t_0$.

\begin{figure}
	\begin{center}
		\resizebox{0.49\textwidth}{!}{\includegraphics{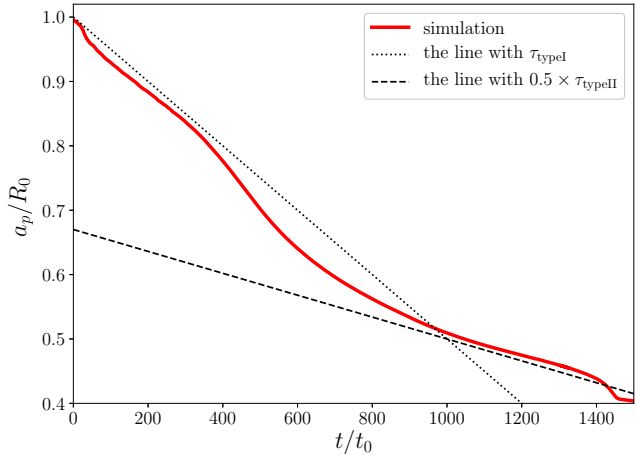}}
		\caption{
		Time variation of planet orbital radius in the case shown in Figure~\ref{fig:dens2d_q1e-4_h5e-2_a3e-4}.
		Thin dashed and dotted lines indicate the lines with the migration velocity of $-5\times 10^{-4} R_0/t_0$ (equivalent to $R_0/\tautypeI$, $\tautypeI$ given by Equation~\ref{eq:tmig_typeI}) and $-1.7\times 10^{-4} R_0/t_0$ ($2R_0/\tauwgap$, $\tauwgap$ given by Equation~\ref{eq:tmig_gap}), respectively.
		\label{fig:rp_q1e-4_h5e-2_a3e-4}
		}
	\end{center}
\end{figure}
Figure~\ref{fig:rp_q1e-4_h5e-2_a3e-4} illustrates the time variation of the planetary orbital radius $\rp$.
Before the planet opens a sufficient gap, the migration timescale may be estimated by that given by type~I migration \citep{Tanaka_Takeuchi_Ward2002,Paardekooper_Baruteau_Crida_Kley2010} as
\begin{align}
\tautypeI & \simeq 2000 \bracketfunc{k}{2}{-1} \bracketfunc{\mpl/\mstar}{10^{-4}}{-1} \nonumber \\
& \qquad  \qquad \times  \bracketfunc{\mstar/(\sigmagas R_0^2)}{10^{3}}{} \bracketfunc{h/R}{0.05}{2} t_0,
\label{eq:tmig_typeI}
\end{align}
where $k$ depends on the disk structure, but for simplicity we adopt $k=2$ in the following.
Note that $t_0=1000$~yr when $R_0=100$~au.
When the planet forms a gap in a stationary state, the migration timescale can be given by \citep{Kanagawa_Tanaka_Szuszkiewicz2018},
\begin{align}
\tauwgap \simeq (1+0.04K)\tautypeI,
\label{eq:tmig_gap}
\end{align}
where $K=(\mpl/\mstar)^2(h/R)^{-5}/\alpha$.
As can be seen in Figure~\ref{fig:rp_q1e-4_h5e-2_a3e-4}, the migration velocity is similar to that given by Equation~(\ref{eq:tmig_typeI}), namely, $R_0/\tautypeI \simeq -5\times 10^{-4} R_0/t_0$.
As the gap opens, the migration velocity slows down, and it becomes $-2\times 10^{-4} R_0/t_0$ around $t=1000 \ t_0$.
This velocity is a factor of two larger than that predicted by Equation~(\ref{eq:tmig_gap}), which is because the migration velocity does not reach the stationary one yet.
Note that we also carried out the simulation with the smaller radius of the inner boundary $R_{\rm in}=0.1 R_0$ (in the fiducial case, $R_{\rm in}=0.3 R_0$), and confirmed that the resulting time variation of $\rp$ is very similar.

\begin{figure}
	\begin{center}
		\resizebox{0.49\textwidth}{!}{\includegraphics{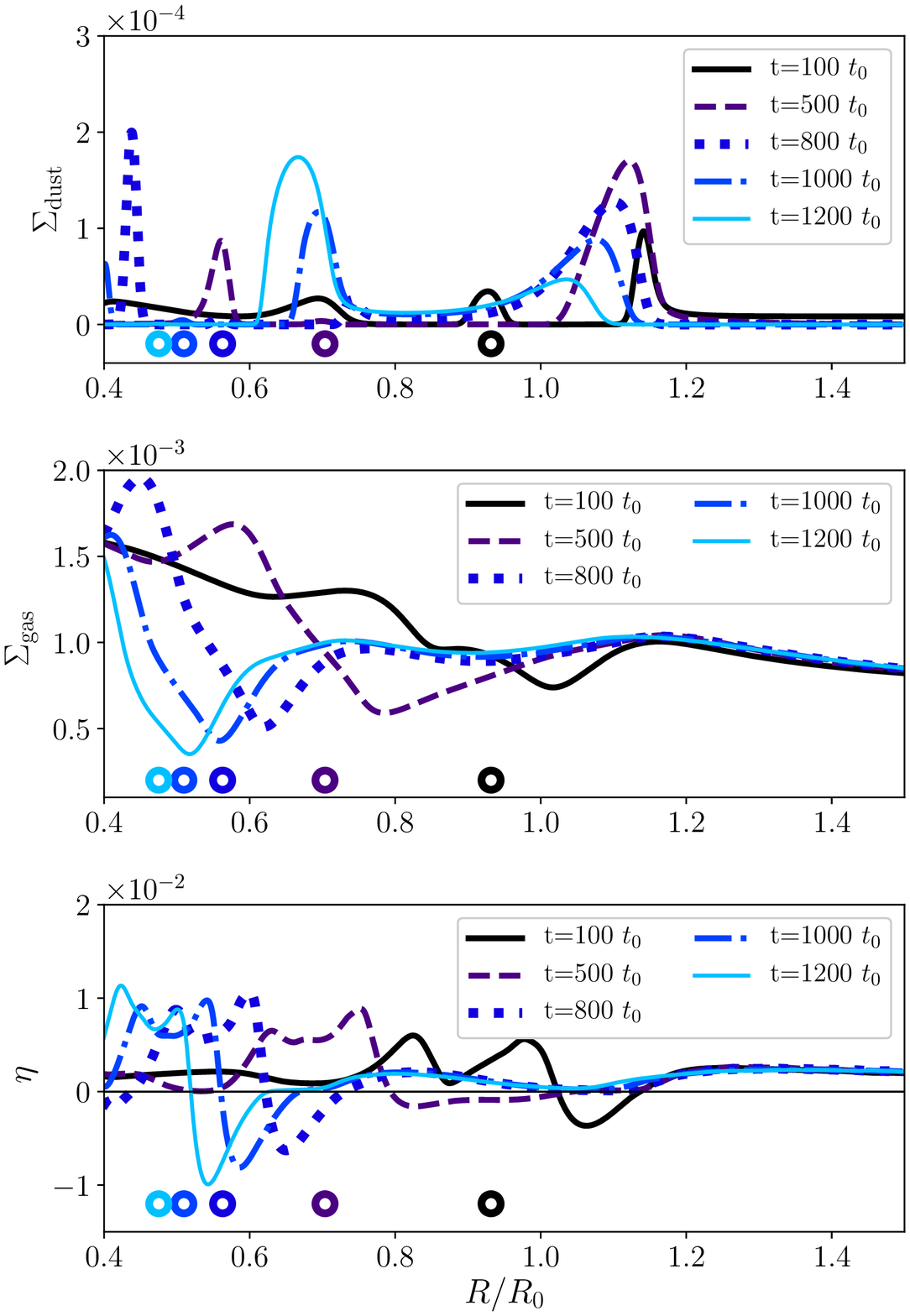}}
		\caption{
		Azimuthally averaged surface density of dust grains (top panel) and gas (middle panel), when $\mpl/\mstar=10^{-4}$, $H_0=0.05$, and $\alpha=3\times 10^{-4}$ (the same as that shown in Figure~\ref{fig:dens2d_q1e-4_h5e-2_a3e-4}).
		The bottom panel shows radial distribution of pressure gradient ($\eta = -1/2 (h/R)^2 \partial \ln P/\partial \ln R$) averaged over azimuth direction.
		The small circles in each panel indicate the location of the planet (which locates at larger radii at an earlier time).
		\label{fig:eta_q1e-4_h5e-2_a3e-4}
		}
	\end{center}
\end{figure}
The formation and deformation of the dust rings are related to the profile of the gas density.
The radial velocity of the dust grains with $\st \ll 1$ can be written by \citep{Nakagawa_Sekiya_Hayashi1986,Kanagawa_Ueda_Muto_Okuzumi2017},
\begin{align}
\vradd &= -2\st \eta \vk,
\label{eq:vrad_dust}
\end{align}
where $\vk$ denotes the Keplerian rotation velocity, and $\eta$ is related to a pressure gradient of the gas defined as
\begin{align}
\eta &= -\frac{1}{2}\bracketfunc{h}{R}{2} \frac{\partial \ln P_{2D}}{\partial \ln R}.
\label{eq:eta}
\end{align}
For convenience, we introduce the drift timescale of the dust grains, $\taudrift \equiv -R/\vradd$, as
\begin{align}
\taudrift &= 800 \bracketfunc{\st}{0.1}{-1} \bracketfunc{\eta}{10^{-3}}{-1} \bracketfunc{R}{R_0}{3/2} t_0.
\label{eq:taudrift}
\end{align}
Note that by comparing the migration velocity of the planet (Equations~\ref{eq:tmig_typeI} and \ref{eq:tmig_gap}), the inward drift of the dust grains is faster in our simulations, but it depends on $\sigmagas$.
The planet can move faster than the dust grains if the larger $\sigmagas$ is adopted.
This issue is further discussed in Section~\ref{subsec:scaling}.

In the upper and middle panels of Figure~\ref{fig:eta_q1e-4_h5e-2_a3e-4}, we illustrate the azimuthal averaged surface densities of dust and gas, respectively, and in the bottom panel of Figure~\ref{fig:eta_q1e-4_h5e-2_a3e-4}, we show the distribution of $\eta$, in the same case shown in Figure~\ref{fig:dens2d_q1e-4_h5e-2_a3e-4}.
At the point that $\eta$ changes from negative to positive from inside, the dust grains are accumulated, which is referred to as the pressure bump (maxima).
As seen in the figure, the pressure bump forms at $R\simeq 1.13 R_0$ at $t=100 \ t_0$, and the dust grains are accumulated around there, which corresponds to the initial dust ring.
The pressure bump associated with the initial ring, which we refer to as the initial pressure bump, does not move while the planet migrates inward until $t\simeq 800 \ t_0$.
In addition to the initial pressure bump, the other pressure bump (the later pressure bump) is formed at $R=0.7R_0$ at $t=800 \ t_0$, which is associated with the gap induced by the planet.
The location of the later pressure bump moves inward as the planet migrates because the planet migrates slowly, and it can reform the gap (also discussed in Section~\ref{subsec:scaling}).

Since the dust feedback makes the gas structure at $\eta=0$, the region with $\eta\simeq 0$ forms within the dust ring, and this dust ring becomes wider as time passes \citep{Kanagawa_Muto_Okuzumi_Taki_Shibaike2018}.
However, the dust feedback does not make $\eta$ positive, because it becomes weak as $\eta$ approaches zero.
The pressure bump is finally deformed by the viscous gas diffusion, instead of the dust feedback, though the feedback changes the structure of the dust ring and its evolution.
Due to the viscous diffusion, $\eta$ becomes positive from the inner edge of the initial ring, and hence the dust grains trapped into the initial pressure bump are gradually released and the initial dust ring is finally deconstructed.
Since the planetary migration is slower than the inward drift of the dust grains, as pointed out above, the released dust grains catch up with the planet.
They are trapped into the inner later pressure bump and form the new dust ring (the later ring).
The initial dust ring completely deforms around $t=1500 \ t_0$, and only the later ring remains after that.

As the pressure bump is formed, the dust mass flux is halted by the pressure bump, and the inward drift of the dust grains resumes when the pressure bump begins to be deformed.
This relation between the dust mass flux and the strength of the pressure bump is discussed later in Section~\ref{subsec:scaling}.
The pressure bump and the initial ring are formed around $R_0$, which identifies the relation between the locations of the planet and rings.
We discuss the physical meaning of $R_0$ in the planet evolution in Section~\ref{subsec:planet_growth}.

\subsection{Dependence on gas viscosity} \label{subsec:depend_gas_viscosity}
\begin{figure*}
	\begin{center}
		\resizebox{0.98\textwidth}{!}{\includegraphics{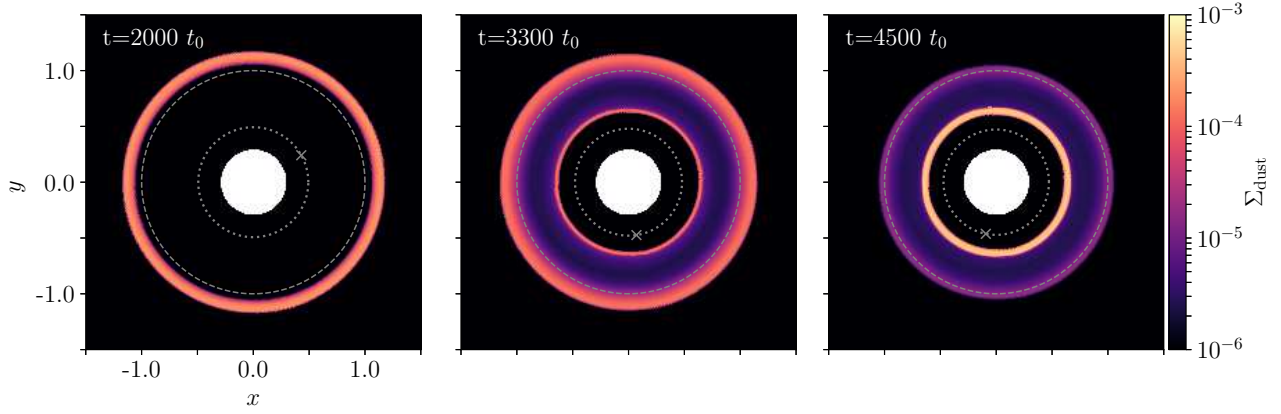}}
		\caption{
		The same as Figure~\ref{fig:dens2d_q1e-4_h5e-2_a3e-4}, but for $\alpha=10^{-4}$.
		\label{fig:dens2d_q1e-4_h5e-2_a1e-4}
		}
	\end{center}
\end{figure*}
With low viscosity, the formation and deformation of the dust rings are similar to the case with $\alpha=3\times 10^{-4}$, but the timescale becomes longer.
Figure~\ref{fig:dens2d_q1e-4_h5e-2_a1e-4} shows the case with $\alpha=10^{-4}$ and where the other parameters are the same as those adopted in the case of $\alpha=3\times 10^{-4}$.
As can be seen in the figure, the dust ring is formed and deformed similarly to the case of $\alpha=3\times 10^{-4}$, but a longer timespan is required to deform the outer initial ring.
This is because the deformation of the pressure bump progresses by viscous diffusion of the gas, and hence, this timescale depends on the viscosity.

\begin{figure}
	\begin{center}
		\resizebox{0.49\textwidth}{!}{\includegraphics{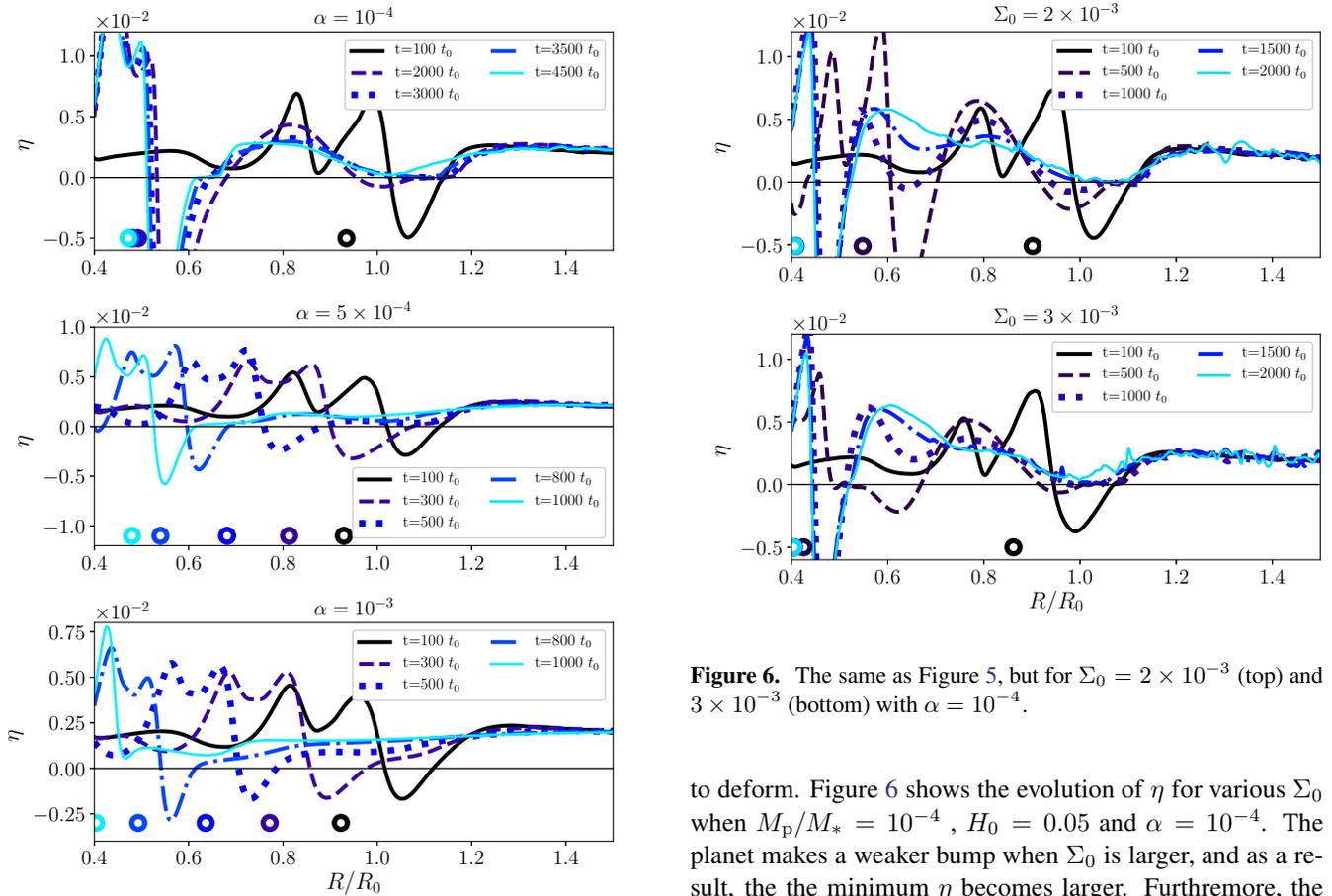}}
		\caption{
		The distributions of $\eta$ in the various $\alpha$, with $\mpl/\mstar=10^{-4}$ and $H_0=0.05$.
		The small circles in each panel indicates are the location of the planet at earlier time (which locates at larger radii at an earlier time).
		\label{fig:eta_q1e-4_h5e-2_avar}
		}
	\end{center}
\end{figure}
The top panel of Figure~\ref{fig:eta_q1e-4_h5e-2_avar} shows the $\eta$ distributions of the gas structure in the case of $\alpha=10^{-4}$ at particular times.
As compared to the case of $\alpha=3\times 10^{-4}$, the minimum $\eta$ is smaller since a stronger bump is formed around the initial location of the planet.
The initial dust ring is formed around the initial pressure bump, and it deforms slowly as time progresses, due to the low viscosity.

The middle and bottom panels of Figure~\ref{fig:eta_q1e-4_h5e-2_avar} show the cases with a relatively large viscosity, namely, $\alpha=5\times 10^{-4}$ and $\alpha=10^{-3}$, respectively.
In the case of $\alpha=5\times 10^{-4}$, the deformation of the initial pressure bump occurs at a similar time as the later pressure bump forms around $t=500 \ t_0$.
In the case of $\alpha=10^{-3}$, the pressure bump is always accompanied by the planet and the dust ring moves inward as the planet migrates, as shown by \cite{Meru_Rosotti_Booth_Nazari_Clarke2018}.
In these two cases, only one dust ring is formed at $R>\rp$.

Note that in the case of $\alpha=10^{-4}$ (the upper panel of Figure~\ref{fig:eta_q1e-4_h5e-2_avar}), the planet reaches close to the inner boundary at $t=1500 \ t_0$.
After that, the planet is trapped by the inner boundary, and its migration stops artificially.
If we adopted a smaller radius of the inner boundary, the planet would migrate further inside (but is difficult due to computational resources).
Although the pressure bump is formed around $R/R_0=0.6$ in the upper panel of Figure~\ref{fig:eta_q1e-4_h5e-2_avar}, these pressure bumps (and later ring as well) should be formed further inside the inner region, as the planet migrates more inward.
Nonetheless, this effect does not change the gas structures around $R=R_0$ because the planet is very distant from that region.
Hence, it does not change the formation and deformation of the initial dust ring, which is the main focus of this study.

\subsection{Dependence on gas surface density} \label{subsec:sigmadep}
\begin{figure}
	\begin{center}
		\resizebox{0.49\textwidth}{!}{\includegraphics{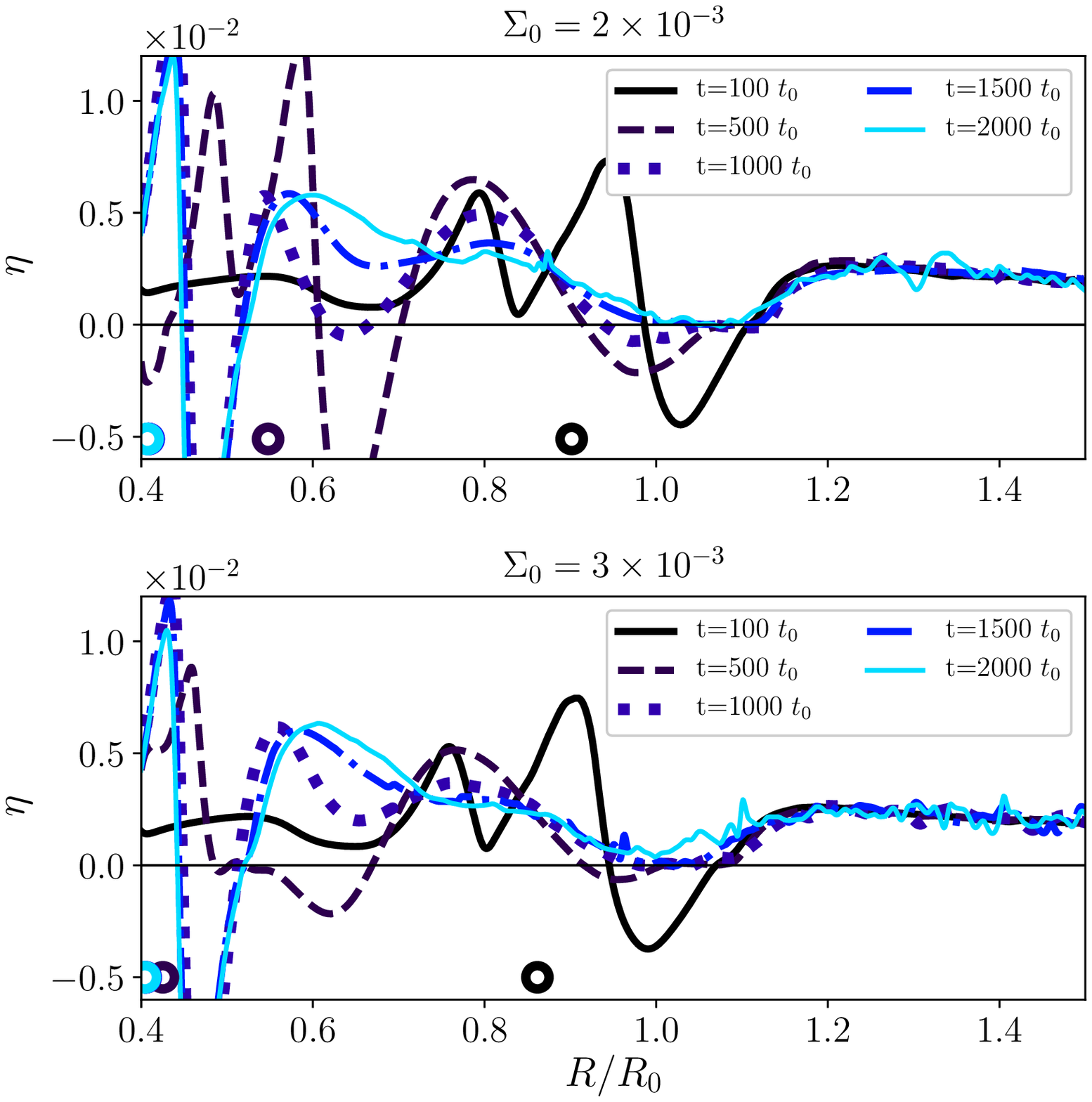}}
		\caption{
		The same as Figure~\ref{fig:eta_q1e-4_h5e-2_avar}, but for $\Sigma_0=2\times 10^{-3}$ (top) and $3\times 10^{-3}$ (bottom) with $\alpha=10^{-4}$.
		\label{fig:eta_q1e-4_h5e-2_a1e-4_Svar}
		}
	\end{center}
\end{figure}
When the planetary migration is faster with a larger $\Sigma_0$, the planet only forms a weak pressure bump at the initial location, which indicates a shorter time for the initial dust ring to deform.
Figure~\ref{fig:eta_q1e-4_h5e-2_a1e-4_Svar} shows the evolution of $\eta$ for various $\Sigma_0$ when $\mpl/\mstar=10^{-4}$ , $H_0=0.05$ and $\alpha=10^{-4}$.
The planet makes a weaker bump when $\Sigma_0$ is larger, and as a result, the the minimum $\eta$ becomes larger.
Furthremore, the deformation timescale of the initial ring becomes shorter as $\Sigma_0$ is larger.
Indeed, the initial pressure bump is completely deformed at $t=2000 \ t_0$ when $\Sigma_0=3\times 10^{-3}$, while it still remains in the case with $\Sigma_0=2\times 10^{-3}$.

Note that in the cases shown in Figure~\ref{fig:eta_q1e-4_h5e-2_a1e-4_Svar}, the planet reaches the inner boundary before the initial dust ring is deformed, as in the upper panel of Figure~\ref{fig:eta_q1e-4_h5e-2_avar}.
As mentioned above, the pressure bumps formed around $R/R_0=0.5$ are formed just due to the artificial stop of the planet migration.

\subsection{Deformation and followability of the dust ring} \label{subsec:scaling}
\begin{deluxetable*}{cccccccccccc}
\tablenum{1}
\tablecaption{Summary of characteristic features \label{tab:charatimes_mineta}}
\tablewidth{0pt}
\tablehead{
\colhead{\#}&\colhead{$\mpl/\mstar$} & \colhead{$H_0$} & \colhead{$\alpha$} & \colhead{$\Sigma_0$} & \colhead{$\etaminovertime$} & \colhead{$\tbumpform/t_0$} & \colhead{$\tbumpformmax/t_0$} & \colhead{$\tstart/t_0$} & \colhead{$\tend/t_0$} & \colhead{$\mbox{max}\left(R_{\rm ring}/\rp \right)$} & \colhead{Feature$^a$}
}
\startdata
1&$10^{-4}$          & $0.05$ & $10^{-3}$          & $10^{-3}$          & $-1.92\times 10^{-3}$ & $7.0 \times 10^{1}$ & $1.7\times 10^{2}$ &$2.8\times 10^{2}$ & $ 8.2\times 10^{2}$ & $1.46$ & follow\\
2&$10^{-4}$          & $0.05$ & $5\times 10^{-4}$  & $10^{-3}$          & $-3.64\times 10^{-3}$ &$6.0 \times 10^{1}$ & $2.0 \times 10^{2}$& $4.1\times 10^{2}$ & $ 9.7\times 10^{2}$ & $1.78$ & left behind\\
3&$10^{-4}$          & $0.05$ & $3 \times 10^{-4}$ & $10^{-3}$          & $-5.22\times 10^{-3}$ & $6.0 \times 10^{1}$& $2.1 \times 10^{2}$& $5.7\times 10^{2}$ & $ 1.5\times 10^{3}$ & $2.08$ & left behind \\
4&$10^{-4}$          & $0.05$ & $10^{-4}$          & $10^{-3}$          & $-8.88\times 10^{-3}$ &$5.0 \times 10^{1}$ &$3.2 \times 10^{2}$ & $2.7\times 10^{3}$ & $ 4.7\times 10^{3}$ & $2.35^b$ & left behind \\
5&$10^{-4}$          & $0.05$ & $10^{-4}$          & $2 \times 10^{-3}$ & $-5.68\times 10^{-3}$ &$4.0 \times 10^{1}$  &$1.8 \times 10^{2}$  & $1.2\times 10^{3}$ & $ 2.6 \times 10^{3}$ & $2.76^b$ & left behind\\
6&$10^{-4}$          & $0.05$ & $10^{-4}$          & $3 \times 10^{-3}$ & $-3.59\times 10^{-3}$ &$4.0 \times 10^{1}$  &$1.0 \times 10^{2}$  & $4.9\times 10^{2}$ & $ 2.0\times 10^{3}$ & $2.70^b$ & left behind \\
\hline
7&$3\times 10^{-5}$ & $0.05$ & $10^{-4}$           & $10^{-3}$          & $\quad 9.68\times 10^{-4}$  & -- & -- & -- & -- & --& no ring\\
8&$5\times 10^{-5}$ & $0.05$ & $10^{-4}$           & $10^{-3}$          & $\quad 1.66 \times 10^{-4}$ &-- & -- & -- & -- & -- & follow \\
9&$6\times 10^{-5}$ & $0.05$ & $10^{-4}$           & $10^{-3}$          & $-1.05\times 10^{-3}$ &$1.3\times 10^{2}$ &$2.9\times 10^{2}$ & $4.7\times 10^{2}$ & $ 1.5\times 10^{3}$ & $1.88$ & left behind \\
10&$6\times 10^{-5}$ & $0.05$ & $3\times 10^{-4}$   & $10^{-3}$          & $-9.84\times 10^{-5}$ &$1.6\times 10^{2} $ & $2.5\times 10^{2}$ & $3.1\times 10^{2}$ & $ 8.2\times 10^{2}$ & $1.53$ & follow \\
11& $6\times 10^{-5}$ & $0.05$ & $5\times 10^{-4}$   & $10^{-3}$          & $\quad 1.96\times 10^{-4}$  & -- & -- & -- & -- & -- & follow \\
\hline
12&$10^{-4}$          & $0.07$ & $10^{-4}$          & $10^{-3}$          & $\quad 7.68\times 10^{-4}$ & -- & -- & -- & -- & -- & no ring\\
13&$2\times 10^{-4}$  & $0.07$ & $10^{-4}$          & $10^{-3}$          & $-1.70\times 10^{-3}$ & $7.0\times 10^{1}$ & $1.9\times 10^{2}$ & $3.9\times 10^{2}$ & $ 1.1\times 10^{3}$ & $2.30$ & left behind \\
14&$3\times 10^{-4}$  & $0.07$ & $10^{-4}$          & $10^{-3}$          & $-7.62\times 10^{-3}$ & $3.0\times 10^{1}$ & $1.6\times 10^{2}$ & $1.2\times 10^{3}$ & $ 2.3 \times 10^{3}$ & $2.23^b$ & left behind \\
15&$2\times 10^{-4}$  & $0.07$ & $3\times 10^{-4}$  & $10^{-3}$          & $-9.27\times 10^{-4}$ & $8.0\times 10^{1} $ & $1.6\times 10^{2}$ & $2.8\times 10^{2}$ & $ 6.0 \times 10^{2}$ & $1.80$ & follow \\
16&$2\times 10^{-4}$  & $0.07$ & $8\times 10^{-5}$  & $10^{-3}$          & $-1.84\times 10^{-3}$ & $7.0\times 10^{1}$ & $2.0\times 10^{2}$ & $4.3 \times 10^{2}$ & $ 1.2 \times 10^{3}$ & $2.54$ & left behind\\
\hline
17&$3\times 10^{-4}$  & $0.10$ & $10^{-4}$          & $10^{-3}$          & $\quad 6.74\times 10^{-4}$ & -- & -- & -- & -- & -- & no ring\\
18&$5\times 10^{-4}$  & $0.10$ & $10^{-4}$          & $10^{-3}$          & $-1.16\times 10^{-4}$ & $1.0\times 10^{2}$ & $2.6\times 10^{2}$ & $4.0\times 10^{2}$ & $ 6.6 \times 10^{2}$ & $2.08$ & left behind \\
19&$6\times 10^{-4}$  & $0.10$ & $10^{-4}$          & $10^{-3}$          & $-2.37\times 10^{-3}$ & $7.0\times 10^{1}$ & $2.3\times 10^{2}$ & $4.0\times 10^{2}$ & $ 8.1\times 10^{2}$ & $2.54$ & left behind \\
20&$8\times 10^{-4}$  & $0.10$ & $10^{-4}$          & $10^{-3}$          & $-6.04\times 10^{-3}$ & $4.0\times 10^{1}$ & $2.3\times 10^{2}$ & $8.8 \times 10^{2}$ & $ 1.3 \times 10^{3}$ & $2.85$ & left behind \\
\enddata
\tablenotetext{a}{Feature of dust ring; (1) no ring, no ring is formed. (2) follow, the dust ring forms aside the planet and follows it. (3) left behind, the initial ring forms and it is left behind by the planet.}
\tablenotetext{b}{the case that the planet reaches close to the inner boundary much earlier than the deformation of the initial dust ring. In this case, $\max(\rring/\rp)$ can be much underestimated.}
\end{deluxetable*}

\subsubsection{Characteristic times of ring deformation}
\begin{figure}
	\begin{center}
		\resizebox{0.49\textwidth}{!}{\includegraphics{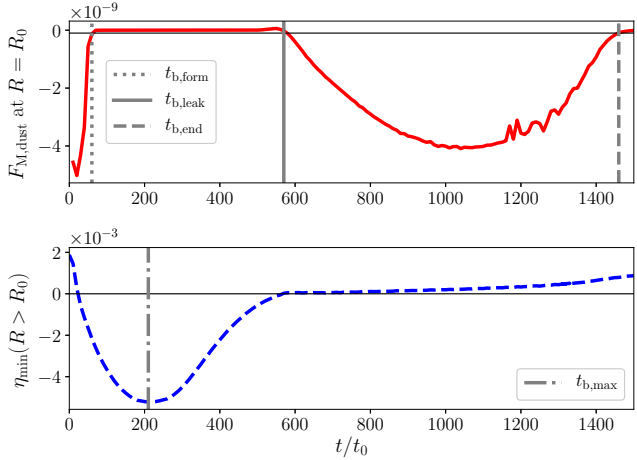}}
		\caption{
		Time variation of the dust flux at $R=R_0$ (top panel) and the minimum value of $\eta$ in the region of $R>R_0$ (bottom panel), when $\mpl/\mstar=10^{-4}$, $H_0=0.05$, and $\alpha=3\times 10^{-4}$.
		\label{fig:epoch_vs_mineta}
		}
	\end{center}
\end{figure}
We first will discuss the timescale of the formation and deformation of the dust rings.
As shown in the previous subsections, the deformation of the initial dust ring and its duration could depend on the viscosity and initial perturbation around the initial location of the planet when the mass of the planet is sufficiently large.
The initial perturbation could be characterized by the value of $\eta$ at $R>R_0$, and the minimum mass of the planet that forms the dust ring can be equivalent to the pebble-isolation mass.
On the basis of this idea, we define two characteristic times associated with the deformation of the initial dust ring by using the dust mass flux at the initial orbital radius of the planet.
In the upper panel of Figure~\ref{fig:epoch_vs_mineta}, we show the time variation of the dust mass flux ($\fmdust$) at $R=R_0$.
As the gap is formed, the dust flux quickly becomes zero at the beginning of the simulation and the initial ring.
As long as $\fmdust=0$ at $R=R_0$, the initial dust ring is retained.
After that, $\fmdust$ starts to decrease as the initial dust ring deforms.
We define two characteristic times, $\tstart$ and $\tend$;
these correspond to the time when $\fmdust$ reaches a threshold value, here we adopt $-10^{-10}$, and $\tstart < \tend$. 
That is, $\tstart$ indicates the time when the dust ring starts to break, and $\tend$ is the time when the dust ring vanishes.
In the bottom panel of Figure~\ref{fig:epoch_vs_mineta}, we show the time variation of the minimum value of $\eta$ in the region of $R>R_0$, which represents the time variation of the initial pressure bump.
In the early phase, $\etamin$ decreases quickly and reaches the minimum value ($=-5.66\times 10^{-3}$) at $t=200 \ t_0$, and then it increases to $0$ as the pressure bump becomes moderate.
In the following, we denote the minimum value of $\etamin$ over the computational time as $\etaminovertime$.
Around $t=\tstart$, $\eta$ reaches $\simeq 0$ and the dust ring gradually deforms, and the dust grains begin to drift inward again.
After $t=\tend$, almost no dust grains remain in the region of $R>R_0$.
The surface density of the dust grains at each time can be seen in Figure~\ref{fig:dens2d_q1e-4_h5e-2_a3e-4}.

Moreover, we define two timescales related to the formation of the initial ring, $\tbumpform$ and $\tbumpformmax$ (they are also denoted in Figure~\ref{fig:epoch_vs_mineta});
the former is the time when $\fmdust$ reaches $-10^{-10}$ at $R=R_0$ for the first time, which indicates the time when the pressure bump is formed.
The latter is the time when $\etamin$ reaches the minimum value ($\etaminovertime$), which corresponds to the time when the pressure bump has been fully developed.

In Table~\ref{tab:charatimes_mineta}, we summarize the $\etaminovertime$, and the characteristic times ($\tbumpform,\tbumpformmax,\tstart,\tend$).
Note that when $\etaminovertime>0$, the characteristic times cannot be defined because no pressure bump is formed.
Looking at $\tbumpform$ and $\tbumpformmax$, we found that the initial pressure bump is quickly formed in $t\lesssim 100 \ t_0$, and $\etamin$ has developed by $\sim 200 \ t_0$ in most cases.
These characteristic times do not significantly depend on the planet mass and the disk parameters.
The deformation of the initial pressure bump is associated with $\tstart$ and $\tend$.

\subsubsection{Followability of the ring}
\begin{figure}
	\begin{center}
		\resizebox{0.49\textwidth}{!}{\includegraphics{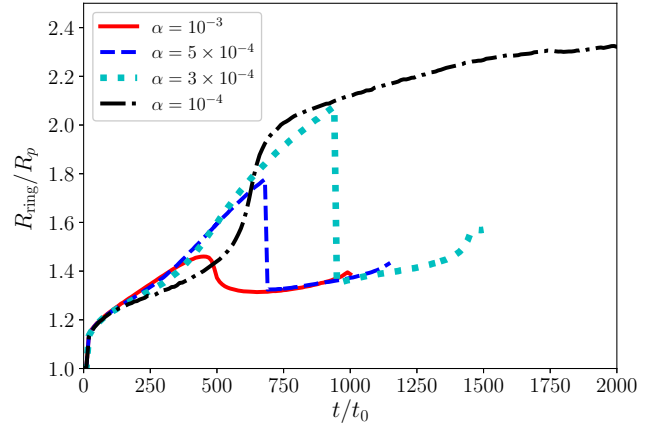}}
		\caption{
		Time variations of the ratio of the location of the dust ring (the maximum peak location of the dust surface density, $\rring$) to the orbital radius of the planet ($\rp$), in the case of $\mpl/\mstar=10^{-4}, H_0=0.05$.
		\label{fig:ratio_rdustring_rp}
		}
	\end{center}
\end{figure}
The followability of the initial dust ring is also an intriguing feature of the structure formed by the migrating planet.
As shown in Section~\ref{subsec:depend_gas_viscosity}, the initial dust ring does not follow the planetary migration in the case with $\alpha \leq 5\times 10^{-4}$, while it moves inward as the planet migrates inward when $\alpha=10^{-3}$.
The ratio of the location of the dust ring to the planetary orbital radius can categorize the two cases, when the ring does/doesn't follow the planet.
Here, we define the location of the dust ring, $\rring$, as the location with the maximum surface density of the dust grains in $R>\rp$.
In Figure~\ref{fig:ratio_rdustring_rp}, we show the ratio of $\rring$ to $\rp$ when $\mpl/\mstar=10^{-4}$, $H_0=0.05$ (the cases shown in Figure~\ref{fig:eta_q1e-4_h5e-2_avar}).
When $\alpha=10^{-3}$, the dust ring follows the planet as can be seen in Figure~\ref{fig:eta_q1e-4_h5e-2_avar}.
In this case, $\rring/\rp$ does not significantly change over time and it is $\sim 1.4$.
However, when the dust ring does not follow the planet (the cases with $\alpha \leq 5\times 10^{-3}$), $\rring/\rp$ increases with time until the deformation of the initial dust ring.
Hence, the maximum value of $\rring/\rp$ during the evolution, $\max(\rring/\rp)$, can be a good indicator of the followability of the dust ring.
Note that a rapid transit of $\rring/\rp$ happens when the peak surface density of the later ring exceeds that of the initial ring, because we define $\rring$ as the location of the maximum surface density in $R>\rp$.
Hence, the rapid transits in Figure~\ref{fig:ratio_rdustring_rp} are just caused by the definition of $\rring$, and are not physical in nature. 

We also summarize $\max(\rring/\rp)$ and features of the dust ring formation in Table~\ref{tab:charatimes_mineta}.
In the cases labeled as 'follow', the dust ring is formed and it follows the planet, and in the cases labeled as 'left behind', the initial dust ring is left behind by the planet.
In the cases labeled as 'no ring', dust grains pass through the planetary orbit and no ring is formed.
One can observe that when $\max(\rring/\rp) \gtrsim 1.8$, the initial dust ring no longer follows the planet.
Note that in the cases that the planet reaches close to the inner boundary much earlier than the deformation of the initial ring, the value of $\max(\rring/\rp)$ can be very underestimated (these cases are denoted by superscript $b$ in the table).
In Appendix~\ref{sec:rings_lowmass}, we show several plots of the dust surface density in the cases labeled as 'no ring' and 'follow', though it is shown in Section~\ref{subsec:formation_and_distruction_dustring} in the cases labeled as 'left behind'.

It should be noted that the cases of \#8 and \#11 are labeled as 'follow', though $\etaminovertime>0$ (which means no pressure maxima).
This results because the dust grains are piled up because their inward drift slows down, and the ring structure is formed without the pressure bump, as shown by \cite{Rosotti_Juhasz_Booth_Clarke2016}.
However, in those cases, the dust grains are not trapped in the ring region. 

\subsubsection{Scaling relation of the ring duration time}
\begin{figure}
	\begin{center}
		\resizebox{0.49\textwidth}{!}{\includegraphics{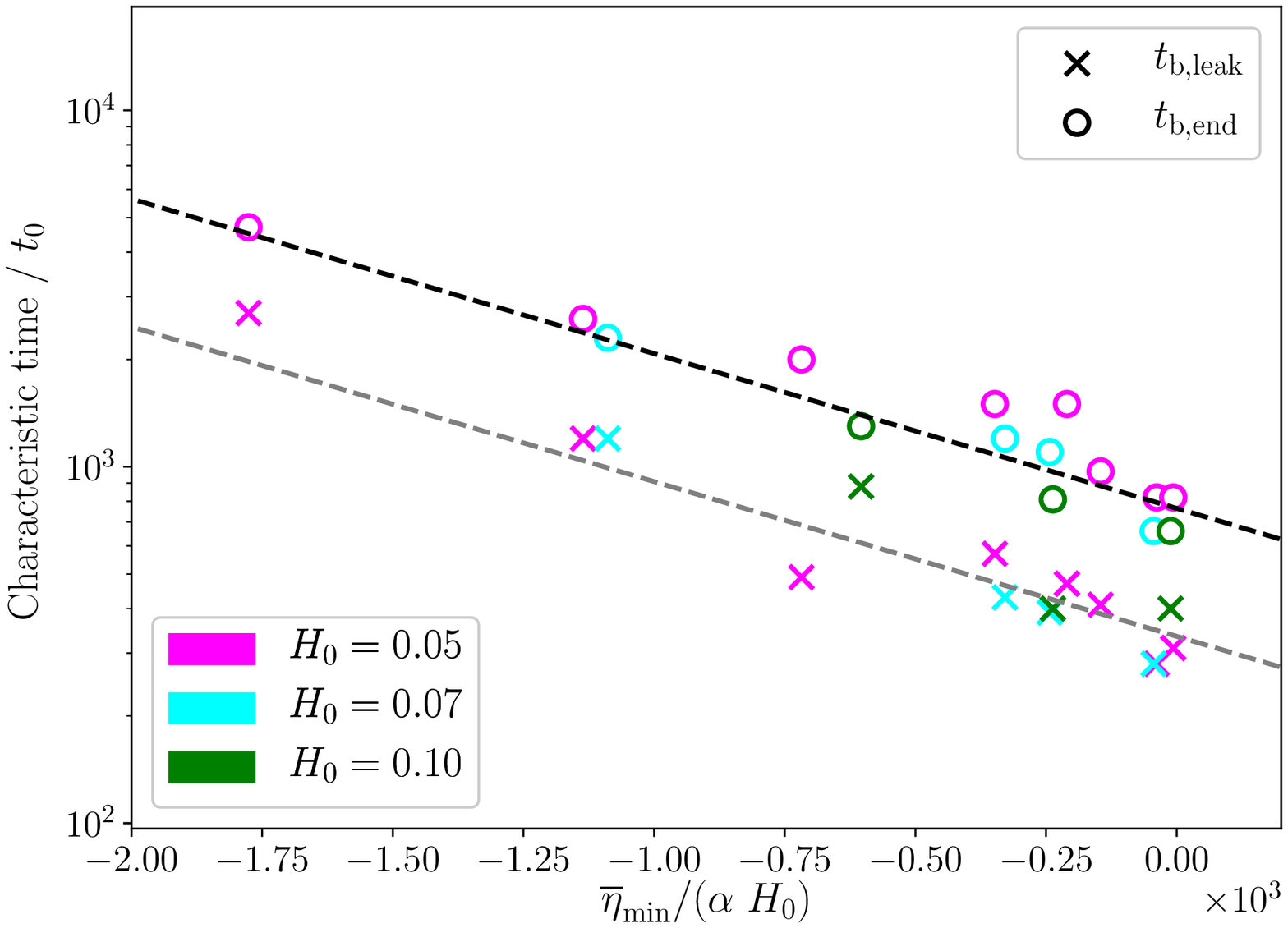}}
		\caption{
		Relation between $\etaminovertime /(\alpha H_0)$ and $\tstart$ and $\tend$ for various setups.
		}
	\end{center}
\end{figure}
As discussed above, $\tstart$ and $\tend$ could depend on the viscosity and $\etaminovertime$.
It might depend on the ratio between the diffusion time, $h^2/\nu$, and the crossing time of the dust grains, $\sim h/\vradd \propto h/(\etaminovertime \vk)$, which $\etaminovertime/(\alpha H_0)$.
On the basis of this idea, we plot the characteristic times, $\tstart$ and $\tend$, and the value of $\etaminovertime/(\alpha H_0)$, in Figure~\ref{fig:epoch_vs_mineta}.
The characteristic times become longer as $\etaminovertime/(\alpha H_0)$ decreases, and we may find the scaling relations for $\tstart$ and $\tend$ against $\etaminovertime/(\alpha H_0)$ as a form of 
\begin{align}
t &= a \exp\left(-0.001 \frac{\etaminovertime}{\alpha H_0} \right) t_0, 
\label{eq:timescales}
\end{align}
where the coefficient $a$ is $3.3\times 10^{2}$ for $\tstart$, and $7.5\times 10^{2}$ for $\tend$.
Note that $t_0=2\pi/\omegak(R_0)$ and, it is 1000~yr when $R_0=100$~au.

The lifetime of the dust ring is closely related to $\etaminovertime$ which can be described by a function of the planet mass, disk scale height, and viscosity, as well as the gas surface density (or migration velocity of the planet).
However, the dependence of these parameters is not clear yet.
Further investigations in wide parameter space are required to build an empirical formula of $\etaminovertime$.

\subsection{Ring evolution for a growing planet} \label{subsec:ring_wacc}
Above, we consider the case with the fixed mass planet by examining the phase before the runaway gas accretion.
After the onset of the runaway gas accretion, the mass of the planet increases rapidly due to the gas accretion, which could affect the ring structure.
In this section, we carried out the simulations with the gas-accreting planet and investigated the effect of the gas accretion on the ring structures.
Instead of using the detailed model of the gas accretion, we adopted the simple model developed by \cite{Kley1999} with the accretion efficiency $\facc$, which is described in Section~\ref{sec:model}.

\begin{figure}
	\begin{center}
		\resizebox{0.49\textwidth}{!}{\includegraphics{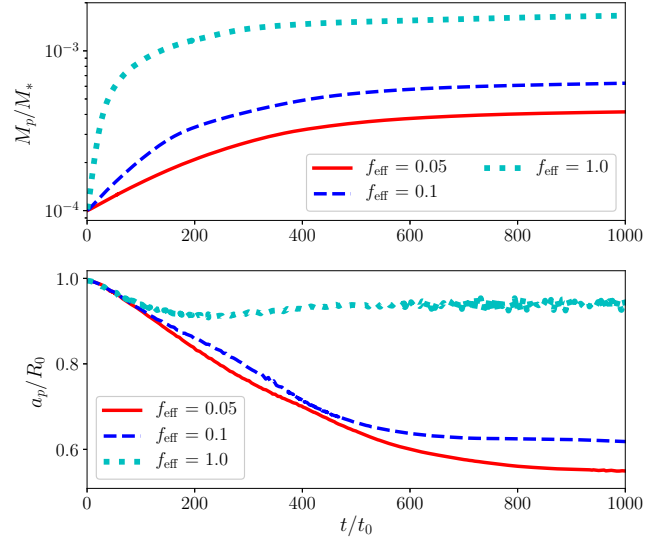}}
		\caption{
		Time variations of the mass (top) and orbital radius (bottom) of the planet with various accretion efficiencies.
		\label{fig:evo_pmass_orbit}
		}
	\end{center}
\end{figure}
We set $\mpl/\mstar=10^{-4}$ as the initial mass of the planet with $H_0=0.07$ and $\alpha=10^{-4}$ (this initial mass is smaller than the mass of the planet that forms the pressure bump and hence $\etaminovertime>0$ in the case with no mass growth, as shown in Table~\ref{tab:charatimes_mineta}).
We adopted three accretion efficiencies, $\facc=0.05$, $0.1$, and $1$.
In Figure~\ref{fig:evo_pmass_orbit}, we show time variations of the planet mass and orbital radius in the cases with the different $\facc$.
As $\facc$ increases, the mass of the planet increases, and the inward migration velocity of the planet becomes slow.
At $t=1000 \ t_0$, $\mpl/\mstar \simeq 4\times 10^{-4}$ in the case with $\facc=0.05$, and $\mpl/\mstar \simeq 6\times 10^{-4}$ and $1.6\times 10^{-3}$ in the cases with $\facc=0.1$ and $\facc=1$, respectively.
The inward migration of the planet suddenly slows down around $R \sim 0.6R_0$ in the cases of $\facc=0.05$ and $0.1$, which is due to the inner boundary.
However, it does affect the deformation of the dust ring.
In the case of $\facc=1$, the migration of the planet is stalled around $R=0.9 R_0$, which can be seen in the case with the low viscosity \citep[e.g.,][]{Kanagawa2019}.
As can be seen in Figure~1 of \cite{Kanagawa2019}, this stalled migration is just transient and the planet migrates inward after a while ($t>2000 \ t_0$).

\begin{figure*}
	\begin{center}
		\resizebox{0.98\textwidth}{!}{\includegraphics{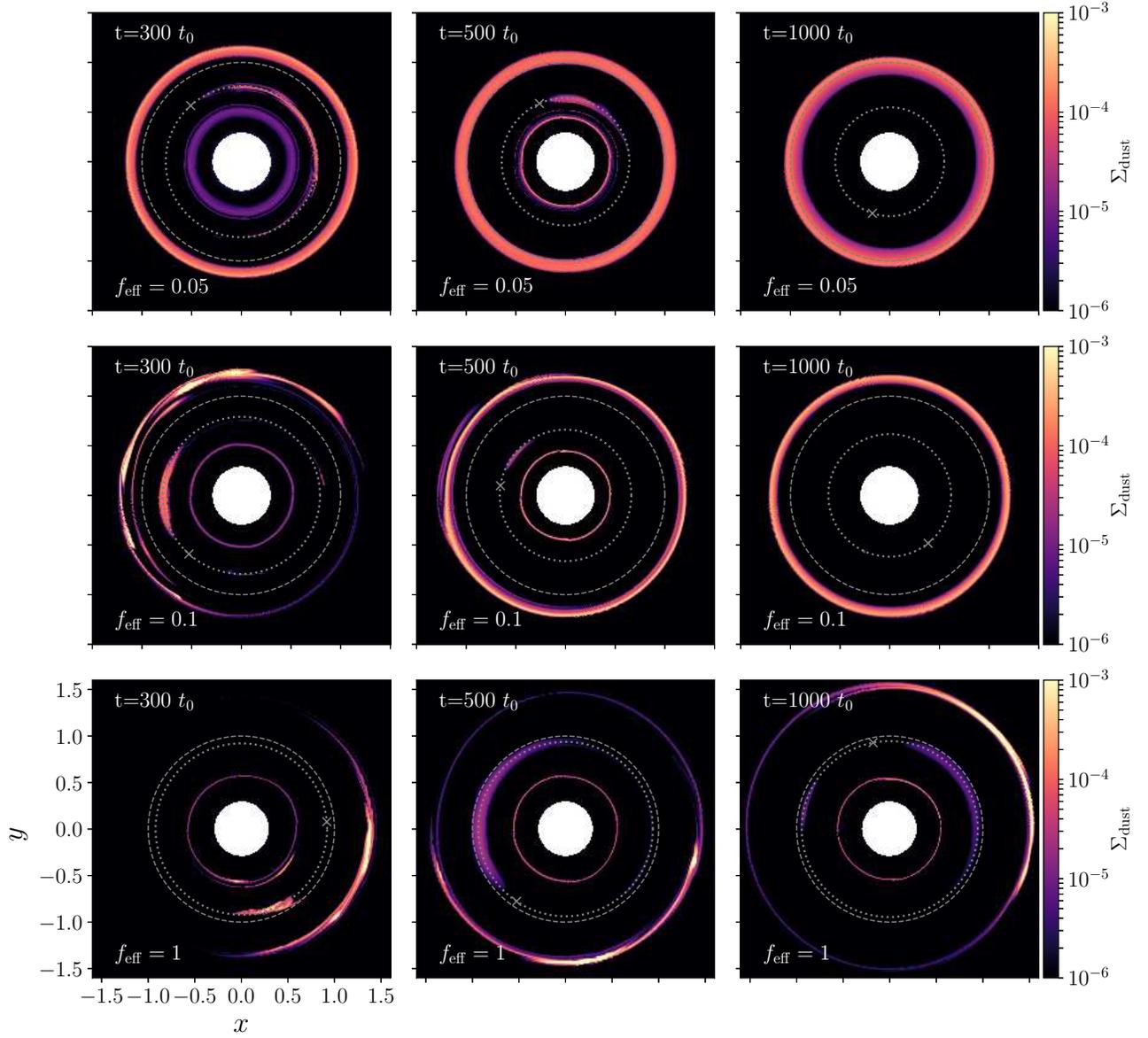}}
		\caption{
		Two-dimensional distributions of the dust surface densities for $\facc=0.05$, (top panels), $0.1$ (middle panels), and $1$ (bottom panels) for the specific times.
		The dashed and dotted circles indicate $R_0$ and $\rp$ and the cross indicates the planet position, as the same as these shown in Figure~\ref{fig:dens2d_q1e-4_h5e-2_a3e-4}.
		\label{fig:dens2d_q1e-4_h7e-2_a1e-4_acc}
		}
	\end{center}
\end{figure*}
We show the surface density of the dust grains in the cases of $\facc=0.05$, $0.1$ and $1$ in Figure~\ref{fig:dens2d_q1e-4_h7e-2_a1e-4_acc}.
For $\facc=0.05$, the formation of the dust ring is quite similar to the case with no-accretion, and the asymmetric ring is formed.
In the case of $\facc=0.1$, the dust ring formed around $R_0$ has several peaks and complex structures at $t=300\ t_0$.
These complex structures become weaker with time, and the ring is almost symmetric and similar to the case with no-accretion, at $t=1000\ t_0$.
When $\facc=1$, a full ring structure is not formed and a semicircular incomplete ring is formed at $t<1000 \ t_0$.
Moreover, in the early phase of $<500 \ t_0$, two or more high duct concentrated peaks are formed on the ring.
In this case, the dust grains can remain for a longer time at the coorbital radius of the planet.
Note that though we did not calculate until the initial ring formed around $R_0$ is deformed, it will be gradually deformed as shown in Figure~\ref{fig:dens2d_q1e-4_h5e-2_a3e-4} with further calculation.

\begin{figure}
	\begin{center}
		\resizebox{0.49\textwidth}{!}{\includegraphics{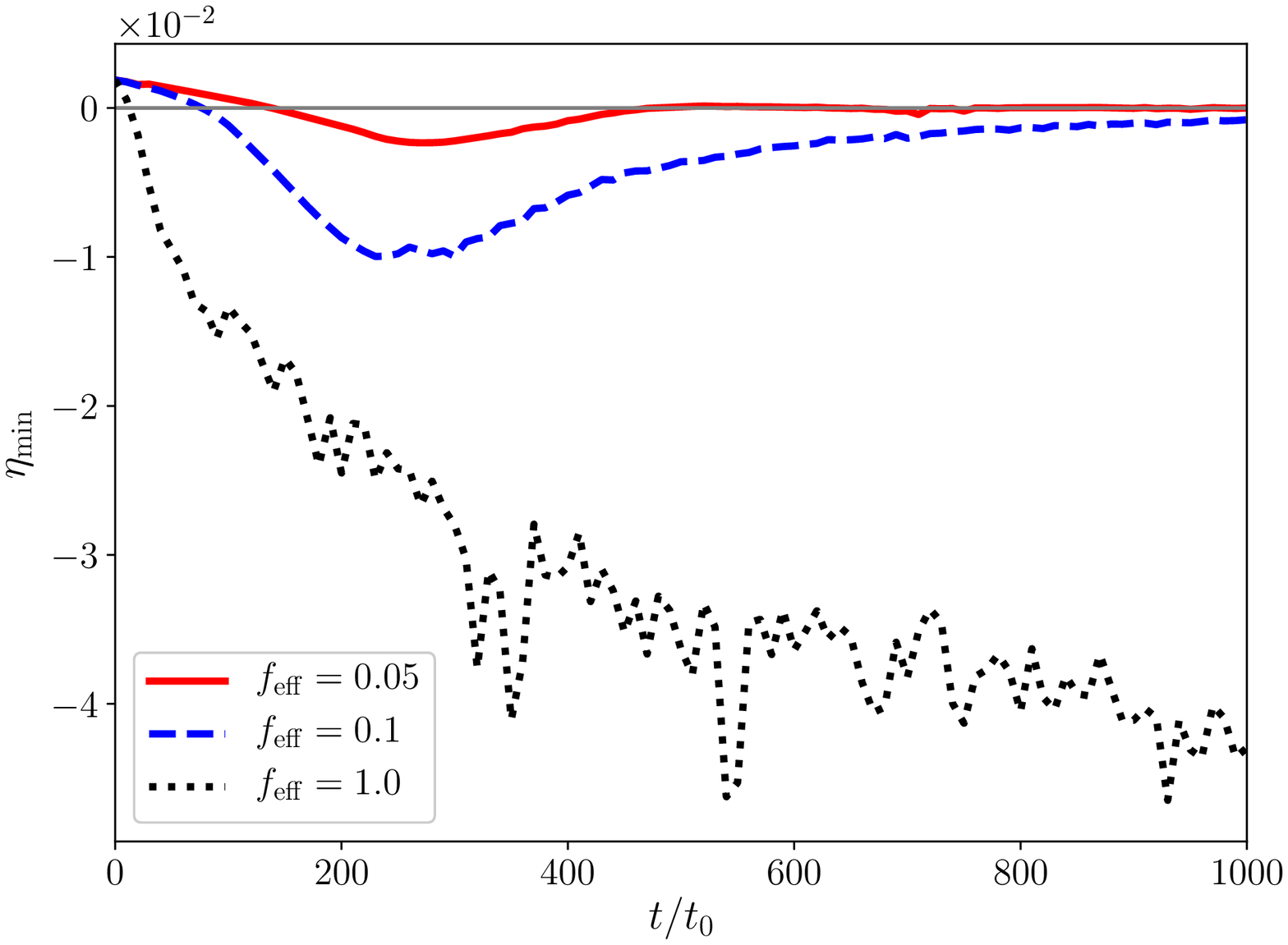}}
		\caption{
		Time variation of $\etamin$ in the case with gas accretion for $\facc=0.05$, $0.1$, and $1$.
		\label{fig:mineta_acc}
		}
	\end{center}
\end{figure}
Figure~\ref{fig:mineta_acc} shows the time variation of $\eta$ with $\facc=0.05$, $0.1$, and $1$.
The variations of $\etamin$ in the cases with $\facc=0.05$ and $0.1$ are similar to those with no-accreting planet shown in the upper panel of Figure~\ref{fig:eta_q1e-4_h5e-2_a3e-4}.
The value of $\etaminovertime$ is $-2.34\times 10^{-3}$ in the case of $\facc=0.05$, and it is $-9.97\times 10^{-3}$ in the case of $\facc=0.1$.
However, for $\facc=1$, $\etamin$ decreases slowly and reaches the minimum value $-4.45\times 10^{-2}$, because the planetary migration slows.
It implies that the asymmetric structures of the dust grains can survive a long time in the case of $\facc=1$.

\begin{figure*}
	\begin{center}
		\resizebox{0.98\textwidth}{!}{\includegraphics{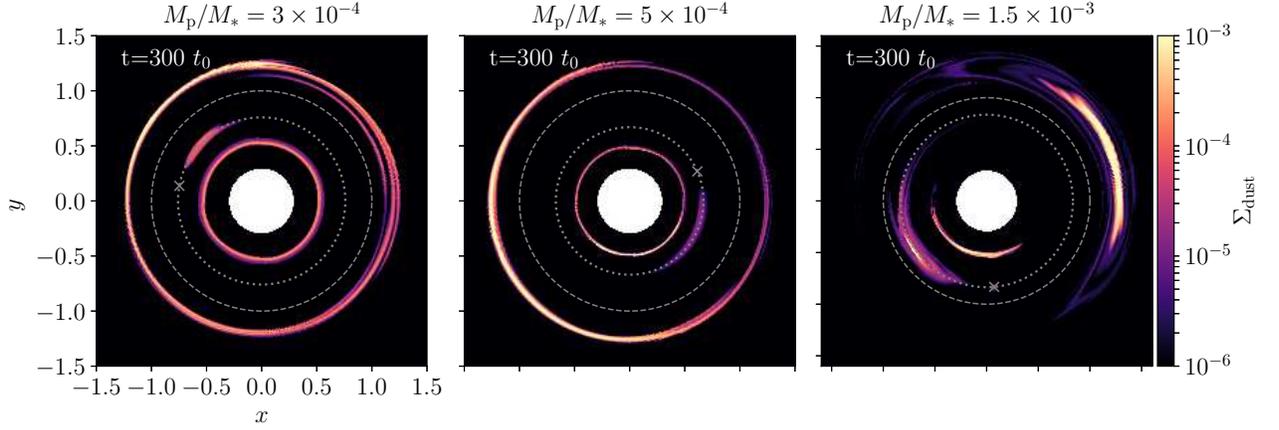}}
		\caption{
		Two-dimensional dust surface distributions with the fixed mass of the planet, $\mpl/\mstar=3\times 10^{-4}$, $\mpl/\mstar=5\times 10^{-4}$, and $1.5\times 10^{-3}$ at $t=300 \ t_0$, for comparison to the results of the cases with the gas accretion.
		The dashed and dotted circles indicate $R_0$ and $\rp$ and the cross represents the location of the planet, as the same as these shown in Figure~\ref{fig:dens2d_q1e-4_h7e-2_a1e-4_acc}.
		\label{fig:dens2d_q1e-4_h7e-2_a1e-4_noacc}
		}
	\end{center}
\end{figure*}
A planet as massive as Jupiter can form a vortex and asymmetric ring structure, as reported by \cite{Lyra_Johansen_Klahr_Piskunov2009}.
To see the difference between the structures formed by the fixed mass planet and by the planet with the gas accretion, we carried out additional simulations with the fixed mass of $\mpl/\mstar=5\times 10^{-4}$ and $\mpl/\mstar=1.5\times 10^{-3}$, keeping the other parameters the same as those adopted in the simulations with the gas accretion.
In Figure~\ref{fig:dens2d_q1e-4_h7e-2_a1e-4_noacc}, we show the dust surface density at $t=300 \ t_0$ in the case of $\mpl/\mstar=3\times 10^{-4}$, $5\times 10^{-4}$, and $1.5\times 10^{-4}$.
Since the mass at $t=300 \ t_0$ in the case with the gas accretion is $\mpl/\mstar=2\times 10^{-4}$ for $\facc=0.05$, $\mpl/\mstar=4\times 10^{-4}$ for $\facc=0.1$, and $\mpl/\mstar=1.2\times 10^{-3}$ for $\facc=1$, the cases shown in Figure~\ref{fig:dens2d_q1e-4_h7e-2_a1e-4_noacc} are good comparisons between the fixed mass case and the gas accretion case.
Similar to that shown in Figure~\ref{fig:dens2d_q1e-4_h7e-2_a1e-4_acc}, some complex and symmetric structures are formed in the ring when $\mpl/\mstar=3\times 10^{-4}$ and $\mpl/\mstar=5\times 10^{-4}$.
Hence, the complex ring structure shown in Figure~\ref{fig:dens2d_q1e-4_h7e-2_a1e-4_acc} could have originated because the mass increases to a certain level.
When $\mpl/\mstar=1.5\times 10^{-3}$, however, the prominent vortex structures are seen in the inner and outer parts of the planetary orbit.
In this case, the vortex-like structures can be found in the gas structure, similar to the dust structure.
The similar vortexes induced by the Rossby wave instability \citep[RWI,]{Li_etal2000,Ono2016,Ono2018} are reported by \cite{Dong_Li_Chiang_Li2018}.
When $\facc=1$, on the other hand, no significant vortex can be seen and only the incomplete ring is formed.
Similarly, no vortex structure is found in the gas structure.
It may imply that though the incomplete ring could be related to the RWI, the formation of the prominent vortex may be prevented by the mass growth of the planet in the outer disk of the planetary orbit.
In the inner disk of the planetary orbit, only a symmetric ring is formed in the case with the mass-growing planet, though in the case with the fixed-mass planet, the prominent vortex is formed by the RWI.
Note that in the simulations carried out by the previous studies \citep{Lyra_Johansen_Klahr_Piskunov2009,Dong_Li_Chiang_Li2018}, the dust grains are accumulated around both the Lagrange points $L_4$ and $L_5$, when the mass and orbital radius of the planet are fixed.
However, in our simulations with the migrating and mass-growing planet, the dust grains are concentrated only around $L_5$ point.
This difference of the dust concentration along the co-orbital radius of the planet could give a clue of the planetary migration if these clumpy structures are observed in the future.

\section{Discussion} \label{sec:discussion}
\subsection{Condition that a planet leave a dust ring behind} \label{subsec:condition_leaving_dustring}
\begin{deluxetable*}{cccccccccccc}
\tablenum{2}
\tablecaption{Timescales of migration and gap-opening \label{tab:time_mig_gapopening}}
\tablewidth{0pt}
\tablehead{
\colhead{\#}&\colhead{$\mpl/\mstar$} & \colhead{$H_0$} & \colhead{$\alpha$} & \colhead{$\Sigma_0$} & \colhead{$\tautypeI/t_0$} & \colhead{$\tauwgap/t_0$} & \colhead{$\taugapform/t_0$} & \colhead{$\tautypeI/\taugapform$} & \colhead{$\tauwgap/\taugapform$} & \colhead{$M_{\rm p}/M_{\rm iso}$}& \colhead{Feature}
}
\startdata
1&$10^{-4}$          & $0.05$ & $10^{-3}$          & $10^{-3}$          &$2.0\times 10^{3}$ &$4.6\times 10^{3}$ &$7.6\times 10^{2}$ &$2.6$&$6.0$  &1.3 &follow\\ 
2&$10^{-4}$          & $0.05$ & $5\times 10^{-4}$  & $10^{-3}$          &$2.0\times 10^{3}$ &$7.1\times 10^{3}$ &$2.2\times 10^{3}$ &$0.93$&$3.3$ &1.5& left behind\\
3&$10^{-4}$          & $0.05$ & $3 \times 10^{-4}$ & $10^{-3}$          &$2.0\times 10^{3}$ &$1.1\times 10^{4}$ &$4.6\times 10^{3}$ &$0.43$&$2.3$ &1.6& left behind\\
4&$10^{-4}$          & $0.05$ & $10^{-4}$          & $10^{-3}$          &$2.0\times 10^{3}$ &$2.8\times 10^{4}$ &$2.4\times 10^{4}$ &$0.08$&$1.2$  &1.7& left behind\\
5&$10^{-4}$          & $0.05$ & $10^{-4}$          & $2 \times 10^{-3}$ &$1.0\times 10^{3}$ &$1.4\times 10^{4}$ &$2.4\times 10^{4}$ &$0.04$&$0.6$  &1.7& left behind\\
6& $10^{-4}$          & $0.05$ & $10^{-4}$          & $3 \times 10^{-3}$&$6.7\times 10^{2}$ &$9.2\times 10^{3}$ &$2.4\times 10^{4}$ &$0.03$&$0.4$ &1.7& left behind\\
\hline                                                                       
7&$3\times 10^{-5}$ & $0.05$ & $10^{-4}$           & $10^{-3}$          &$6.7\times 10^{3}$ &$1.4\times 10^{4}$ &$7.3\times 10^{3}$ &$0.93$ &$2.0$ &0.52& no ring\\
8&$5\times 10^{-5}$ & $0.05$ & $10^{-4}$           & $10^{-3}$          &$4.0\times 10^{3}$ &$1.7\times 10^{4}$ &$1.2\times 10^{4}$ &$0.33$ &$1.4$ &0.87& follow\\
9&$6\times 10^{-5}$ & $0.05$ & $10^{-4}$           & $10^{-3}$          &$3.3\times 10^{3}$ &$1.9\times 10^{4}$ &$1.4\times 10^{4}$ &$0.23$ &$1.3$ &1.0& left behind\\
10&$6\times 10^{-5}$ & $0.05$ & $3\times 10^{-4}$   & $10^{-3}$          &$3.3\times 10^{3}$ &$8.5\times 10^{3}$ &$2.8\times 10^{3}$ &$1.2$ &$3.1$ &0.95& follow \\
11&$6\times 10^{-5}$ & $0.05$ & $5\times 10^{-4}$   & $10^{-3}$          &$3.3\times 10^{3}$ &$6.4\times 10^{3}$ &$1.3\times 10^{3}$ &$2.6$ &$5.0$ &0.90& follow \\
\hline                                                                                            
12&$10^{-4}$          & $0.07$ & $10^{-4}$          & $10^{-3}$          &$2.8\times 10^{3}$ &$9.5\times 10^{3}$ &$7.4\times 10^{3}$ &$0.38$ &$1.3$ &0.63& no ring\\
13&$2\times 10^{-4}$  & $0.07$ & $10^{-4}$          & $10^{-3}$          &$1.4\times 10^{3}$ &$1.5\times 10^{4}$ &$1.5\times 10^{4}$ &$0.10$ &$1.0$ &1.3& left behind\\
14&$3\times 10^{-4}$  & $0.07$ & $10^{-4}$          & $10^{-3}$          &$9.3\times 10^{2}$ &$2.1\times 10^{4}$ &$2.2\times 10^{4}$ &$0.04$ &$0.94$ &1.9& left behind\\
15&$2\times 10^{-4}$  & $0.07$ & $3\times 10^{-4}$  & $10^{-3}$          &$1.4\times 10^{3}$ &$5.8\times 10^{3}$ &$2.9\times 10^{3}$ &$0.50$ &$2.1$ &1.2& follow \\
16&$2\times 10^{-4}$  & $0.07$ & $8\times 10^{-5}$  & $10^{-3}$          &$1.4\times 10^{3}$ &$1.8\times 10^{4}$ &$2.1\times 10^{4}$ &$0.07$ &$0.87$ &1.3& left behind\\
\hline                                                                                            
17&$3\times 10^{-4}$  & $0.10$ & $10^{-4}$          & $10^{-3}$          &$1.3\times 10^{3}$ &$6.1\times 10^{3}$ &$6.3\times 10^{3}$ &$0.21$ &$0.96$ &0.65& no ring\\
18&$5\times 10^{-4}$  & $0.10$ & $10^{-4}$          & $10^{-3}$          &$8.0\times 10^{2}$ &$8.8\times 10^{3}$ &$1.1\times 10^{4}$ &$0.08$ &$0.83$ &1.1& left behind \\
19&$6\times 10^{-4}$  & $0.10$ & $10^{-4}$          & $10^{-3}$          &$6.7\times 10^{2}$ &$1.0\times 10^{4}$ &$1.3\times 10^{4}$ &$0.05$ &$0.80$  &1.3& left behind\\
20&$8\times 10^{-4}$  & $0.10$ & $10^{-4}$          & $10^{-3}$          &$5.0\times 10^{2}$ &$1.3\times 10^{4}$ &$1.7\times 10^{4}$ &$0.07$ &$0.78$  &1.7& left behind\\
\enddata
\end{deluxetable*}
\begin{figure}
	\begin{center}
		\resizebox{0.49\textwidth}{!}{\includegraphics{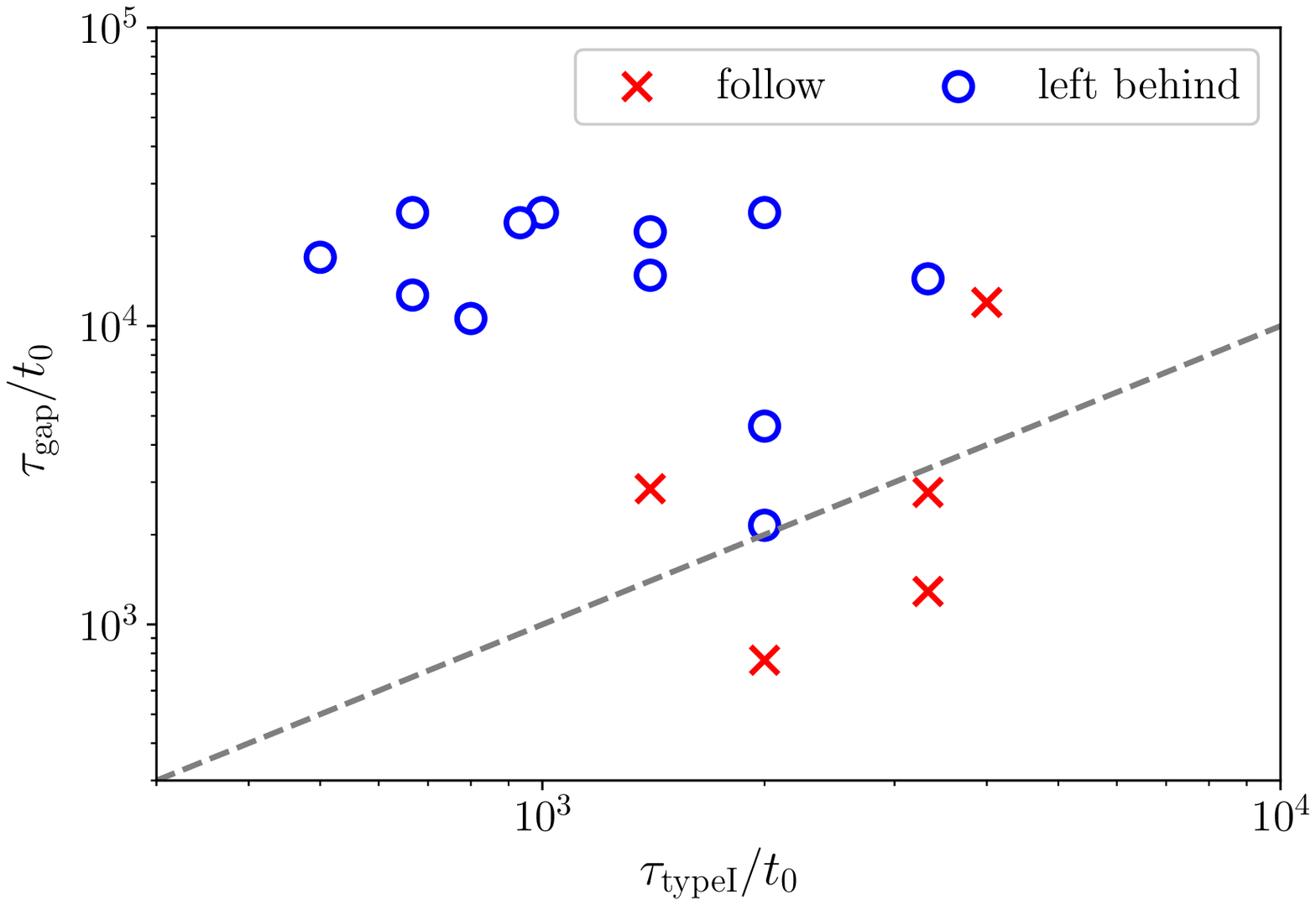}}
		\caption{
		The cases labeled as 'follow' (cross) and the cases labeled as 'left behind' (circle) in the $\tautypeI$ vs. $\taugapform$ plane.
		The dashed declined line indicates the line of $\tautypeI=\taugapform$.
		\label{fig:ttypeI_vs_tgap}
		}
	\end{center}
\end{figure}

The initial ring is formed by the trap of the dust grains in the initial pressure bump.
The pressure bump is associated to be the gap and it vanishes when the gap is closed.
The closing time of the gap can be estimated by the viscous timescale over the width of the gap, $\Delta^2/\nu$, where $\Delta$ is the width of the gap divided by $\rp$ and $\nu$ is the kinetic gas viscosity.
For a stationary gap with a finite gas viscosity, \cite{Kanagawa2017b} found that the formation timescale of the gap can be estimated by $\Delta^2/\nu$, and hence the closing timescale is equivalent to be the formation timescale of the gap  (which we refer to as $\taugapform$).
In this case, $\taugapform$ is given by \citep{Kanagawa2017b}
\begin{align}
\taugapform = 2.4\times 10^{4} \bracketfunc{\mpl/\mstar}{10^{-4}}{} \bracketfunc{h/R}{0.05}{-7/2} \bracketfunc{\alpha}{10^{-4}}{-3/2} t_0.
\label{eq:taugapform}
\end{align}
When $\taugapform$ is shorter than the timescale of the planetary migration, the gap and the pressure bump reform as the planet migrates, because these formation and deformation progress faster than the planetary migration.
Since the migration timescale can be estimated by $\tautypeI$ given by Equation~(\ref{eq:tmig_typeI}) in the early phase, as can be seen in Figure~\ref{fig:rp_q1e-4_h5e-2_a3e-4}, the condition that the dust ring follows the planet can be expressed as $\tautypeI/\taugapform > 1$.
While, the dust ring can be left behind by the planet when $\tautypeI/\taugapform<1$.
In Table~\ref{tab:time_mig_gapopening}, we summarize $\tautypeI$, $\taugapform$, and the ratio of them in each case, with the features that are the same as those listed in Table~\ref{tab:charatimes_mineta}.
One can confirm that the observed features are consistent to the conditions that the dust ring follow/does not follow the planet, described above, in most cases.
For instance, in the case of $\alpha=10^{-3}$, $\mpl/\mstar=10^{-4}$, $H_0=0.05$, and $\Sigma_0=10^{-3}$ (the case of \#1 in the table), $\tautypeI/\taugapform=2.6$ and the pressure bump follows the planet as can be seen in the bottom panel of Figure~\ref{fig:eta_q1e-4_h5e-2_avar}.
In contrast, in the case of \#3, $\tautypeI/\taugapform \lesssim 1$, and the initial pressure bump and the initial ring is left behind by the planet, as shown in Figure~\ref{fig:eta_q1e-4_h5e-2_a3e-4}.

We plot the cases labeled by 'follow' and 'left behind' in Figure~\ref{fig:ttypeI_vs_tgap} (the cases labeled as 'no ring' are not plotted because the initial ring is not formed in those cases).
All the cases labeled 'left behind' are located in the region of $\tautypeI/\taugapform<1$.
In the cases labeled as 'follow', three cases out of five are located in the region of $\tautypeI/\taugapform>1$, though two cases (\#8 and \# 15) are above the line of $\tautypeI/\taugapform=1$.
Since \#15 is one of the most viscous cases in our simulations ($\nu \simeq 1.5\times 10^{-6}$) and only a faint pressure bump is formed ($\etaminovertime\simeq -10^{-3}$), the initial ring deformes quickly and only the last ring is visible for most of the simulation time.
Hence it is labeled as 'follow' (see $\tend$ in Table~\ref{tab:charatimes_mineta}).
The case of \#8 can be a transitional case between 'follow' and 'left behind', because a weak dust ring is formed around $R=R_0$ though $\etaminovertime>0$.
The dust grains are always leaking from the ring and these follow the migrating planet (which is why it is labeled as 'follow').
However, the dust distribution is similar to the case of \#9 (which is 'left behind' case, where $\mpl$ is 1.2 times larger).
The distributions of the dust surface density in the cases of \#8 and \#15 are shown in Appendix~\ref{sec:rings_lowmass}.

The condition of $\tautypeI/\taugapform < 1$ can be rewritten by $\Sigma > \sigmacrit$ and 
\begin{align}
\sigmacrit &= 3.4 \mbox{ g/cm}^2 \bracketfunc{\mstar}{1M_{\odot}}{} \bracketfunc{R_0}{100\mbox{ au}}{-2} \nonumber\\
& \qquad \qquad  \times \bracketfunc{\mpl/\mstar}{10^{-4}}{-2} \bracketfunc{h/R}{0.1}{11/2} \bracketfunc{\alpha}{10^{-4}}{3/2}.
\label{eq:sigmacrit}
\end{align}
If we may roughly estimate the disk mass by $M_{\rm disk} = 2\pi R_0^2 \sigmacrit$ (thus $\sigmagas \propto R/R_0$ and a disk with a radius of $R_0$ is assumed), the disk mass corresponding to $\sigmacrit$ can be written by
\begin{align}
\frac{M_{\rm disk,crit}}{\mstar} &= 0.024 \bracketfunc{\mpl/\mstar}{10^{-4}}{-2} \bracketfunc{h/R}{0.1}{11/2} \bracketfunc{\alpha}{10^{-4}}{3/2}.
\label{eq:mdcrit}
\end{align}
Hence, $\mdiskcrit$ is close to the disk mass of the standard minimum solar nebula model ($= 0.01 \mstar$) when $\alpha \simeq 10^{-4}$, while it is much larger than the standard mass when $\alpha=10^{-3}$.
When $M_{\rm disk} > M_{\rm disk,crit}$, the planet migration can be faster than the gap formation (thus $\tautypeI/\taugapform < 1$) in the early phase and the initial dust ring will be left behind by the planet.
The value of $M_{\rm disk,crit}$ highly depends on aspect ratio and viscosity, and it becomes small with smaller $h/R$ and $\alpha$.
Hence, the location of the dust ring can be largely different from the location of the planet, in a cooler disk with lower viscosity even in a lighter disk.

The migration velocity of the planet slows down as time passes, as shown in Figure~\ref{fig:rp_q1e-4_h5e-2_a3e-4}.
In the inner region, the migration timescale of the planet can be estimated by $\tauwgap$.
The ratio of $\tauwgap$ to $\taugapform$ can indicate the followability of the later ring in the inner region; that is, when $\taugapform/\taugapform \gtrsim 1$, the later ring can follow the migrating planet.
Hence we also show $\tauwgap/\taugapform$ in Table~\ref{tab:time_mig_gapopening}.
As can be seen in Table~\ref{tab:time_mig_gapopening}, $\tauwgap/\taugapform \gtrsim 1$ in most of the cases, and hence the later ring follows the migrating planet in our simulations.
But for \#5 and \#6, $\tauwgap/\taugapform << 1$ due to the larger $\Sigma_0$.
In these cases, even in the inner region, the location of the planet can be much different from that of the later ring, though this effect is not visible in our simulations due to the inner boundary.

As can be seen in Table~\ref{tab:time_mig_gapopening}, the planet migration is slower than the inward drift of the dust grains estimated by Equation~(\ref{eq:taudrift}), $\taudrift \sim 800 (\st/0.1)^{-1} \ t_0$ in our setup of the simulations.
However, in the case with the larger $\Sigma_0$, the planet migration can be faster than the inward drift of the dust grains.
Alternatively, when the size of the dust grains is smaller, the drift velocity of the dust grains is slower than the migration velocity of the planet, because it becomes slow with smaller dust grains.
In these cases, while the initial dust ring is deformed, the dust grains leaking from the initial ring cannot catch up with the planet, and hence the later ring cannot be formed.
Alternatively, if the migration timescale is much shorter than $\tend$, the planet will fall to the central star before the dust ring is deformed.
In this case, the later ring will also not be formed.

It requires a sufficient large mass of the planet to form the dust ring.
We may estimate this sufficient mass by the pebble-isolation mass.
The pebble-isolation mass,$\miso$, can be written by \citep{Bitsch_Morbidelli_Johansen_Lega_Lambrechts_Crida2018} 
\begin{align}
\frac{\miso}{\mstar} &\simeq 7.5\times 10^{-5} \bracketfunc{\hp}{\rp}{3} \left[0.34 \bracketfunc{-3}{\log\left(\alpha\right)}{4}+0.66\right].
\label{eq:miso}
\end{align}
We describe the ratio of $\mpl$ to $\miso$ in Table~\ref{tab:time_mig_gapopening}.
In all the 'no ring' cases, one can find that $\mpl/\miso<1$.
The planet mass is larger than the pebble isolation mass in all the 'left behind' cases.
In three 'follow' cases (\#8, \#10, and \#11), the dust ring is formed with $\mpl/\miso \simeq 0.9$.
Although there are a few exceptions in the 'follow' case, the pebble-isolation mass can be regarded as the minimum mass for the 'left behind' case.

\subsection{Connection to observed ring structure} \label{subsec:obs_rings}
\begin{figure*}
	\begin{center}
		\resizebox{0.98\textwidth}{!}{\includegraphics{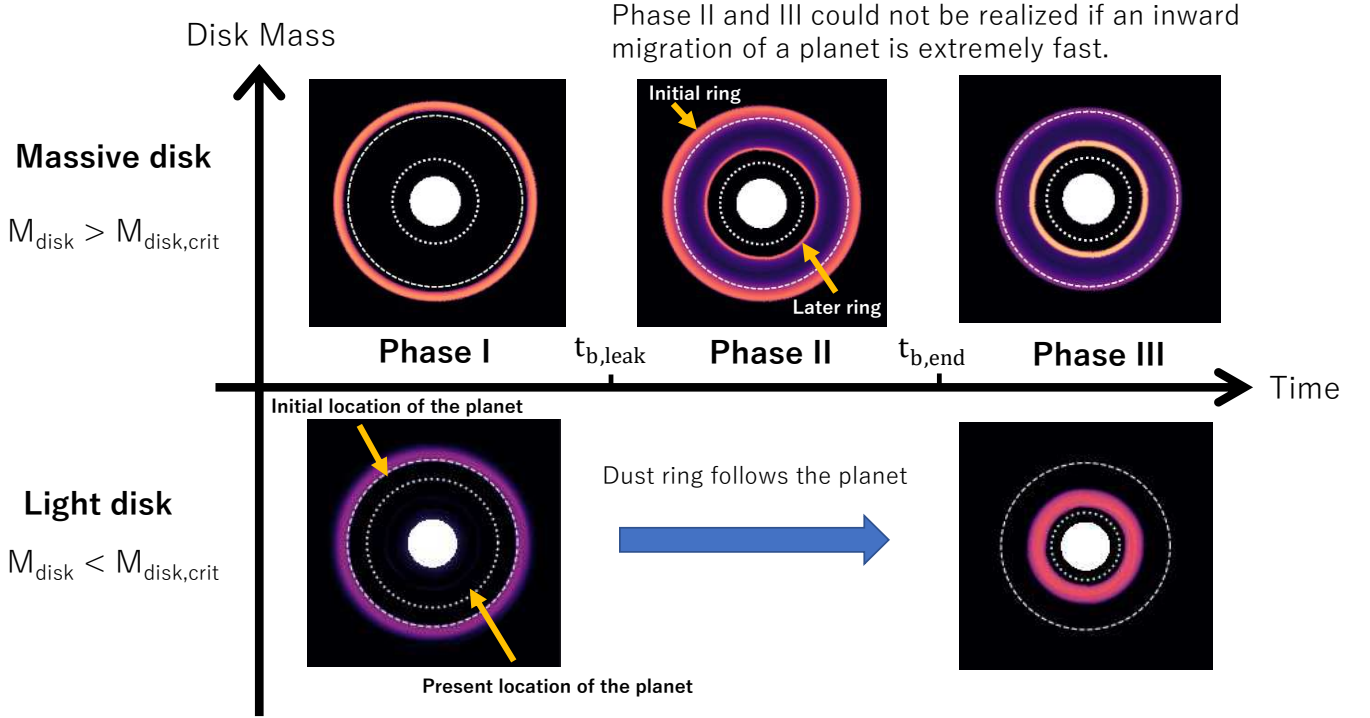}}
		\caption{
		Schematic picture of the morphology of the dust rings and definition of evolutionary phases.
		As stated in the text, when the migration timescale of the planet is much shorter than dust drift timescale and deformation time of the initial (outer) ring, Phase II and III are not realized.
		\label{fig:phase}
		}
	\end{center}
\end{figure*}
As discussed in the previous subsection, when $\sigmagas < \sigmacrit$ given by Equation~(\ref{eq:sigmacrit}) or $\mdisk < \mdiskcrit$ (Equation~\ref{eq:mdcrit}), the dust ring follows the planet, and hence the location of the dust ring is always associated the location of the planet, which is consistent with the results of \cite{Meru_Rosotti_Booth_Nazari_Clarke2018}.
On the other hand, when $\sigmagas > \sigmacrit$ or $\mdisk > \mdiskcrit$, the the dust ring does not follow the migrating planet.
When $\alpha = 10^{-3}$, $\mdiskcrit= 0.76\mstar$, which is much larger than the typical mass of a protoplanetary disk, namely, $\mdisk \sim 0.01\mstar$, and hence, the dust ring likely follows the planet.
While, when $\alpha=10^{-4}$ and $h/R=0.1$, $\mdiskcrit= 0.024 \mstar$, which is comparable to the typical mass of the disk.
Hence, in the case with $\alpha \lesssim 10^{-4}$, the location of the dust ring can be much different from that of the planet.
In this case, the migrating planet can make characteristic ring structures, depending on the time, as shown in Section~\ref{sec:results}.
First, the initial ring is formed around the initial location of the planet, and it does not move from the initial location.
We call Phase I the phase that only the initial dust ring exits, namely $t<\tstart$.
As time passes, the initial dust ring is gradually deformed.
Hence, the dust grains trapped into the initial ring drift inward again, and these are trapped again into the later ring formed close to the present position of the planet.
We call this phase in which two dust rings exist ($\tstart<t<\tend$) Phase II.
After the initial ring disappears ($t>\tend$), only the later ring exists, and we call the phase Phase III.
As discussed in Section~\ref{subsec:condition_leaving_dustring}, Phase II and III may not be realized when the planet migration is much faster than the radial drift of the dust grains and deformation of the initial dust ring.
We summarize the above classification in Figure~\ref{fig:phase}, and the observed morphology of the dust rings may tell us the history of the planet evolution as its footprint.
In this subsection, we discuss what we can learn from these properties of the dust rings and how to relate to the observed rings, when $\mdisk > \mdiskcrit$ and $\mpl \gtrsim \miso$.

When $t<\tstart$ (Phase I), the location of the dust ring indicates the location where the planet was formed, alternatively the location where the mass of the planet reaches the mass to form a sufficient pressure bump (it could be around the pebble-isolation mass), rather than the present position of the planet.
From an observational point of the view, many dust rings located a distance from a central star, $\sim 100 $~au, are observed, as reported by e.g., \cite{Long2018,DSHARP1,VanderMarel2019}, and the turbulent viscosity in these disks may be low \citep[e.g.,][]{Flaherty2015,Teague2016TWHydra,Pinte2016}.
The timescale of this phase can be estimated by $\tstart$ listed in Table~\ref{tab:charatimes_mineta}, or estimated by Equation~(\ref{eq:timescales}).
If $R_0=100$~au, $\mstar=1\ M_{\odot}$, and hence $t_0=1000$~yr, $\tstart \gtrsim 1$~Myr, which is comparable to the typical disk lifetime, 1~Myr \citep{Haisch_Lada_Lada2001}.
Thus, this phase could be observable.
Hence, the observed distant rings might represent the birthplace of the protoplanet, which is further discussed in Section~\ref{subsec:planet_growth}.

In the range from $\tstart$ to $\tend$ (Phase II), the initial ring and the later ring co-exist at $R>\rp$, as can be seen in e.g., Figure~\ref{fig:dens2d_q1e-4_h5e-2_a3e-4}.
The disks with two high-contrast rings at the outer region, such as the disk of AS~209, HD~163296, and HD~143006, could correspond to this phase.
Moreover, a plateau region of the dust emission outside the ring as seen in the disk of Elias~24 can be naturally produced by the migrating planet, as seen in the snapshot at $t=1500 \ t_0$ of Figure~\ref{fig:dens2d_q1e-4_h5e-2_a3e-4}, and the snapshot at $t=4500 \ t_0$ of Figure~\ref{fig:dens2d_q1e-4_h5e-2_a1e-4}.
Though such a structure is hard to be explained by other mechanisms, e.g., snow line, sintering-effect, the secular gravitational instability, etc.
According to Equation~(\ref{eq:timescales}), a duration time of this phase can be written by $\tend-\tstart = 4.14\times 10^{2} \exp\left(-0.001 \etaminovertime/\alpha \right) t_0$.
Intriguingly, the ratio of the duration times of Phase I and II can be independent of $\alpha$ and $\etaminovertime$, and it is given by $\sim 1$.
This implies that the observational probability of the disk with a single distant ring (the disk in Phase I) is similar to that of the disk with double rings (the disk in Phase II) if these are formed by the fast migrating planet.
However, it should exclude a very massive disk, because Phase II never appears in the massive disk because of the very fast planet migration.

Moreover, in Phase~II the spatial interval between the inner and outer rings is related to the migration velocity of the planet, which depends on the disk properties, namely, viscosity, scale height and gas density, and the history of the planet mass growth.
For instance, in the case of Elias~24, the stellar age is $\sim 0.2$~Myr and two rings are observed at 123~au and 77~au (and the gaps are also identified at 89~au and 57~au) by the DSHARP \citep{DSHARP1,Huang_DSHARP}.
Hence, it is reasonable to assume that the initial and current positions of the planet are $\sim 100$~au and $\sim 60$~au, and hence the migration velocity can be estimated by $\sim 200$~au/Myr and the migration timescale is $\sim 0.5$~Myr.
\cite{Zhang_DSHARP2018} estimated the mass of the planet within the gap at 57~au as $\sim 0.4 M_J$ when $\alpha=10^{-4}$ (the estimated mass depends on the property of the dust grains, and hence we adopted the value with their 'DSD1' dust model).
The actual migration timescale can be estimated by the velocity between that given by the type~I regime ($\tautypeI$, Equation~\ref{eq:tmig_typeI}) and that given by the type~II regime ($\tauwgap$, Equation~\ref{eq:tmig_gap}), as shown in Figure~\ref{fig:rp_q1e-4_h5e-2_a3e-4}.
When adopting $\mpl/\mstar=5\times 10^{-4}$, $h/R=0.1$, $\alpha=10^{-4}$, and $R_0=100$~au, we can estimate $\tautypeI$ as $1.8 (1\mbox{ g/cm}^2/\sigmagas)$~Myr and $\tauwgap=18 (1\mbox {g/cm}^2/\sigmagas)$~Myr.
By comparing the migration timescale estimated by the gap separation, $\sim 0.5$~Myr, the gas density at $100$~au can be estimated by $\sim 4 \mbox{ g/cm}^2$ for $\tautypeI$ and $\sim 40 \mbox{ g/cm}^2$ for $\tauwgap$.
These estimated $\sigmagas$ are roughly consistent to that estimated by \cite{Zhang_DSHARP2018} (30~g/cm$^2$) and the upper limit of the gas density estimated by \cite{DuLlemond_DSHARP2018} (18~g/cm$^2$ at 77~au).
In the case of AS~209, the outer two rings are observed at 74~au and 120~au \citep{Huang_DSHARP}, and the stellar age is about 1~Myr \citep{DSHARP1}.
Hence, we can assume that the initial and current positions of the planet are $100$~au and $60$~au, and hence the estimated migration velocity is $\sim 40$~au/Myr and the migration timescale is $\sim 2.5$~Myr.
When $\mpl=0.3 \ M_J$ and the same parameters as the case of Elias~24 are adopted, $\tautypeI\simeq 3$~Myr and $\tauwgap \simeq 14$~Myr when $\sigmagas=1 \mbox{ g/cm}^2$.
Therefore, the estimated $\sigmagas$ is $ \sim 1 \mbox{g/cm}^2$ for $\tautypeI$ and $\sim 6 \mbox{ g/cm}^2$ for $\tauwgap$, which is also consistent with the estimate of \cite{DuLlemond_DSHARP2018} (0.26~g/cm$^2$ to 6.9~g/cm$^2$ at 120~au).
In the case of HD~143006, two rings are observed at 42~au and 65~au, and \cite{Zhang_DSHARP2018} estimated the planet within the gap at $22$~au as $10 M_J$.
In the case of such a large planet, the fixed mass of the planet during the migration may not be appropriate because its migration velocity quickly slows down as the gap is formed.
To explain the observed separation, we need to assume a very large gas density such that the disk becomes gravitationally unstable.
It may indicate that these rings are not associated with the migrating planet, or, it may imply a significant mass growth of the planet during its migration, as shown in Section~\ref{subsec:ring_wacc}.
The asymmetric structure of the outer ring is observed in the disk of HD~143006, which may be related to the asymmetric structure due to the rapid mass growth shown in the bottom panel of Figure~\ref{fig:dens2d_q1e-4_h7e-2_a1e-4_acc}.
More sophisticated modeling with hydrodynamic simulations will improve the above estimate and constrain not only planet mass and disk property, but also the accretion and migration history of the planet.

After $\tend$ (Phase III), only the later ring can be observed and the initial ring completely vanishes.
In this phase, most dust grains are trapped into the planet-associate pressure bump, and the size of the dust disk shrinks as the planet migrates inward.
This phase may correspond to a small disk of the dust grains which have been observed by the recent survey \citep{ODISEAI}.
According to the theory of the viscous accretion disk model, the lifetime of the smaller disks is shorter \citep{Pringle1981}, and hence, the smaller disk should be rare to observe.
However, if the inward drift of the grains is interpreted by the planet, many small dust disks can be observed because the lifetime is determined by the planetary migration.
In this case, the cavity or the gap structure could be observed in the unresolved region, and the gas disk should be larger than the dust disk, which will be revealed by future high-angular resolution observations.

Although many dust rings are observed at the outer region of the protoplanetary disks, attempts to obtain direct imaging of the giant planets in the outer region of the protoplanetary disk \citep[e.g.,][]{Zurlo2020} have yielded a few successes such as PDS~70 b and c \citep{PDS70_Keppler2018,PDS70_Muller2018,Haffert2019}.
This absence of the giant planets may be related to their initial conditions \citep{Brittain_Najita_Dong_Zhu2020}.
Alternatively, it may be due to the inward migration of the planet.
In this case, the giant planet could be detected in the inner region of the disk.

\subsection{Dust trap and debris disks} \label{subsec:debri_disk}
Recent surveys of protoplanetary disks have revealed that the relation between the mass of the dust disk and the stellar mass for the transitional disks is less steep than that for other disks \citep[e.g.,][]{Ansdell2017,Pinilla2018,vanderMarel_Mulders2021}.
This indicates that the transitional disks tend to have a larger dust mass than that of non-transitional disks, with a given stellar mass.
This trend could be due to the dust trap by the pressure bump \citep[e.g.,][]{Pinilla_Birnstiel_Ricci_Dullemond_Uribe_Testi_Natta2012,Pinilla2020}.
If an amount of the dust grains are held in the pressure bump at the outer region, the lifetime of the dust disk becomes longer.
\cite{Michel_vanderMarel_Matthews2021} suggest a longer lifetime of the disk ($\sim 7$ -- $8$~Myr), by considering the disk fraction in the older stellar forming regions, such as $\eta$~Cha, TW~Hya, and Upper~Sco.
The dust trap may be the origin of the debris disks, as discussed by \cite{Michel_vanderMarel_Matthews2021}.
A planet can create the pressure bump, and as shown in this paper, the timescale of the dust trap can be estimated by $\tend$ given by Equation~(\ref{eq:timescales}).
To sustain the dust ring for $\sim 10$~Myr, $\etaminovertime \simeq 1.6 \times 10^{-2}$ is required when $\alpha=10^{-4}$, $H_0=0.07$ at $R=100$~au from $\tend$ given by Equation~(\ref{eq:timescales}).
Such a pressure bump can be formed if $\mpl/\mstar > 3\times 10^{-4}$ (see Table~\ref{tab:charatimes_mineta}, $\etaminovertime=-7.6\times 10^{-3}$), or it can be formed by the gas-accreting planet with an $\facc$ which is between $0.1$ and $1$ (see Figure~\ref{fig:mineta_acc}).
In any case, a relatively large planet might be located in the inner region of the long-lived disk.

\subsection{Implications for planet formation} \label{subsec:planet_growth}
The initial ring is formed at the initial location of the planet, $R_0$, which is one of the intriguing locations of structures induced by the migrating planet shown in this paper.
In reality, the initial location $R_0$ indicates the location where the planet becomes large enough to create a sufficient pressure bump and prominent dust ring, though it is just an input parameter in our simulations.
If the planet increases size by pebble accretion, this initial location can be regarded as the point at which the planet reaches close to the pebble-isolation mass.
In this case, depending on the duration to the onset of the runaway gas accretion, the evolution is different.
If the duration is relatively long, the cases with the fixed mass planets, as shown in Section~\ref{subsec:formation_and_distruction_dustring}, could be appropriate.
Though the case with the gas accretion, as shown in Section~\ref{subsec:ring_wacc}, could be preferred when the runaway gas accretion starts quickly.
Alternatively, a relatively large planet may be formed by the fragmentation of the dust ring induced by the secular gravitational instability \citep{Takahashi_Inutsuka2014,Takahashi_Inutsuka2016a}.
In this case, the location that the instability is caused corresponds to the location of $R_0$.
If the core is formed close to the water snowline and it moves to the outer region by the scattering, this scattered planet can grow up to $0.1$ -- $10 M_J$ until its orbit becomes circular \citep{Kikuchi_HIguchi_Ida2014}.
The location that the orbit of the scattered planet is circular corresponds to $R_0$.
By the gravitational fragment of the massive disk, the giant planet can be formed within a short time, and it falls to the central star rapidly \citep{Baruteau_Meru_Paardekooper2011}.
This location that the fragmentation occurs is also the candidate of our initial location of $R_0$.

Once the planet with sufficient mass is formed at $R_0$, it will migrate inward due to the disk-planet interaction.
Since the planet migrates inward rapidly, the ring location can be much different from that of the planet, and the planet is located at a much smaller radii.
The two high-contrast rings observed in e.g., AS~209, HD~163296, and HD~143006, could indicate the initial and the present locations of the planet, from the outer, respectively, as discussed in Section~\ref{subsec:obs_rings}. 
It may be consistent with the fact that no signature of planets close to the rings is detected except a few systems, while these signatures are easier to detect at larger radii.
Moreover, recent observation of TW~Hya done by \cite{Nomura2021} marginally detect the difference between the gap center (deepest location of the gap) and the location of the planet-like point source, which can be formed by the rapid inward migrating planet \citep{Kanagawa_Nomura_Tsukagoshi_Muto_Kawabe2020}.
This observation may indicate that the planet forms outside and migrates inward.
If so, the planet could be observed at the inner region of the disk, even if the dust ring is observed at the outer region of the disk.

When the gas accretion starts, the ring structure can deviate from the symmetric, as shown in Section~\ref{subsec:ring_wacc}.
As the mass of the planet increases, this incomplete ring with the high-concentrated peaks is significant, and it can remain for a long time, namely, $t=1000 \ t_0$.
These incomplete rings can be related to those observed in the disks of HD~143006, HD~34282, and HD~100453.
On the other hand, a large vortex may be difficult to form by the mass-growing planet.
A large vortex observed in e.g., the disk of HD~142527, may be formed by the Rossby wave instability \citep[e.g.,][]{Li_etal2000,Ono2016,Ono2018}, or the initially massive planet that is formed by e.g., gravitational fragmentation of the massive disk.

\section{Summary} \label{sec:summary}
In this study, we have investigated the formation and deformation of dust rings induced by a rapidly inward migrating planet in disks with a relatively low viscosity, by carrying out gas--dust two-fluid hydrodynamic simulations.
Our results are summarized as:
\begin{enumerate}
  \item We found that a dust ring induced by a planet does not follow the migrating planet for the relatively massive disks which cause rapid planetary migrations, while a dust ring can always follow the planet in the less massive disk.
  The critical disk mass dividing the two cases, $\mdiskcrit$, is given by Equation~(\ref{eq:mdcrit}) and is typically $\sim 0.02 \mstar$ for cases with a relatively low viscosity of $\alpha \sim 10^{-4}$.
  \item  When $\mdisk > \mdiskcrit$, as shown in Figure~\ref{fig:dens2d_q1e-4_h5e-2_a3e-4}, the location and the number of the gaps vary according to the migration of the planet, which can be categorized into three phases as summarized in Figure~\ref{fig:phase}.
  In the early phase (Phase I), a single dust ring (the initial ring) forms at the right outside of the initial formation site of the planet. The initial ring does not follow the migrating planet but remains at the initial location.
  In the next phase (Phase II), the initial ring is gradually deformed and the the dust grains leak from the initial ring and drift inward.
  The leaking dust grains are trapped in the outside of the gap created by the migrating planet and a new ring (the later ring) forms.
  These two rings co-exist until the initial one disappears completely. After that, only the later ring remains outside the planet (Phase III).
  \item  The evolution of the ring morphology is characterized by two timescales on the pressure bump, $\tstart$ and $\tend$.
  Phase~II starts at $\tstart$ and ends at $\tend$.
  These timescales are given by Equation~(\ref{eq:timescales}) (see also Figure~\ref{fig:epoch_vs_mineta}).
  \item The durations of Phase~I ($\tstart$) and Phase~II ($\tend-\tstart$) are comparable and estimated to be $\sim 1$~Myr, when the planet is migrating from $R_0=100$~au.
  Hence these phase are sufficiently long to be observed.
  \item We also investigated the effect of the mass accretion of the planet on the ring formation (Section~\ref{subsec:ring_wacc}). 
  As the mass of the planet increases, a semicircular incomplete ring can be formed instead of the symmetric complete ring. These structures can be related to the asymmetric structures observed in e.g., the disks of HD~143006 and HD~34282.
  However, the formation of a large vortex may be interfered with by the mass-growing planet. 
\end{enumerate}
Our simulation data is available on Zenodo \dataset[doi:10.5281/zenodo.5509281]{https://doi.org/10.5281/zenodo.5509281}.

\section*{}
We would like to thank Dr. Ruobing Dong for his careful review and constructive comments that are helpful to improve the manuscript.
KDK was supported by the JSPS KAKENHI grant  19K14779.
Numerical computations were carried out on the Cray XC50 at the Center for Computational Astrophysics, National Astronomical Observatory of Japan.
\software{FARGO \citep{Masset2000}, Matplotlib \citep[\url{http://matplotlib.org}]{Matplotlib}, NumPy \citep[\url{http://www.numpy.org}]{NumPy}}

\appendix
\section{Effects of numerical setup} \label{sec:effect_numerical_setup}
\subsection{Initial mass growth}
\begin{figure}
	\begin{center}
		\resizebox{0.49\textwidth}{!}{\includegraphics{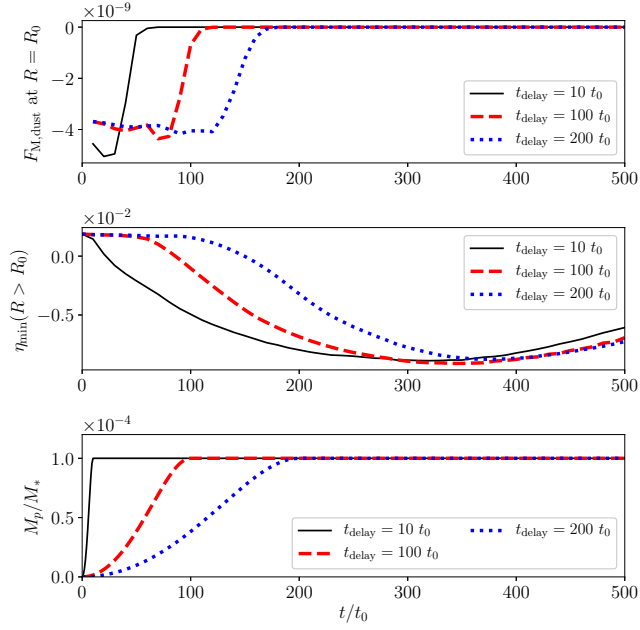}}
		\caption{
		Time variations of the dust flux at $R=R_0$ (top panel), the minimum value of $\eta$ in the region of $R>R_0$ (middle), and the mass of the planet (bottom), in the cases with $\tdelay=10 \ t_0$ (fiducial), $t=100 \ t_0$, and $t = 200 \ t_0$.
		The aspect ratio, and the $\alpha$-parameter are $0.05$ and $10^{-4}$, respectively.
		\label{fig:flux_MTaper}
		}
	\end{center}
\end{figure}
Here we discuss the effect of our numerical setup on the results.
As described in Section~\ref{sec:model}, we adopted gradual mass growth of the planet at the beginning of the simulation, to avoid the abrupt entry of the planet.
The mass of the planet increases as $\mpl \sin\left[\pi t^2/\left(2 \tdelay^2 \right)\right]$ in $t<\tdelay$, and it reaches the given value of $\mpl$ at $t=\tdelay$.
In the fiducial simulations shown in Section~\ref{sec:results}, $\tdelay=10 \ t_0$ is adopted.
In Figure~\ref{fig:flux_MTaper}, we show $\fmdust$ at $R=R_0$, $\etamin$ in the region of $R>R_0$, and the mass of the planet, for the cases with $\tdelay=100 \ t_0$, $200 \ t_0$, and the fiducial value of $10 \ t_0$.
With the longer $\tdelay$, the formation of the pressure bump is delayed, but this delay time is almost comparable to $\tdelay$.
The value of $\etaminovertime$ is also very similar in the cases with the different $\tdelay$.
Hence, we conclude that the choice of $\tdelay$ does not significantly change our results. 

\subsection{Inner boundary}
\begin{figure*}
	\begin{center}
		\resizebox{0.98\textwidth}{!}{\includegraphics{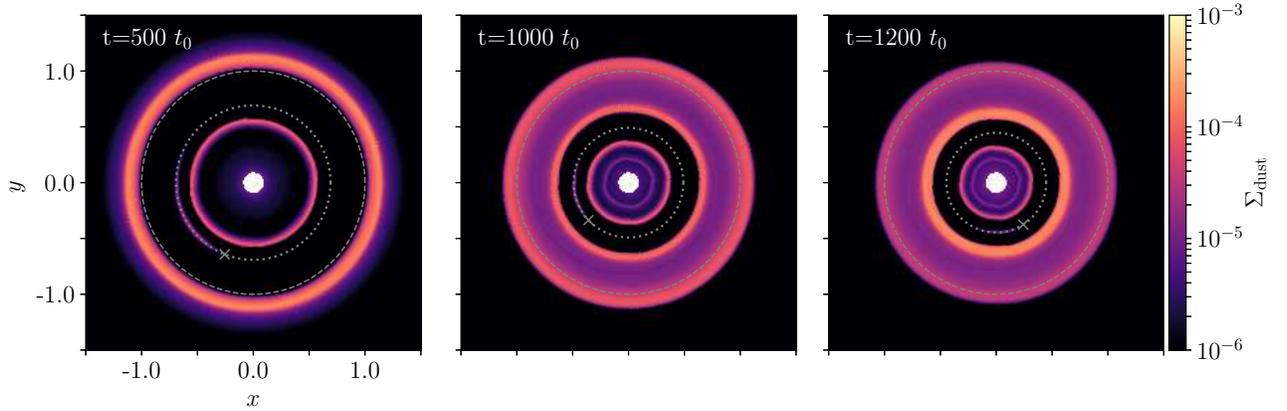}}
		\caption{
		The same as Figure~\ref{fig:dens2d_q1e-4_h5e-2_a3e-4}, but the case with the smaller radius of the inner boundary ($R_{\rm in}=0.1 R_0$).
		\label{fig:dens2d_rin0.1}
		}
	\end{center}
\end{figure*}
\begin{figure}
	\begin{center}
		\resizebox{0.49\textwidth}{!}{\includegraphics{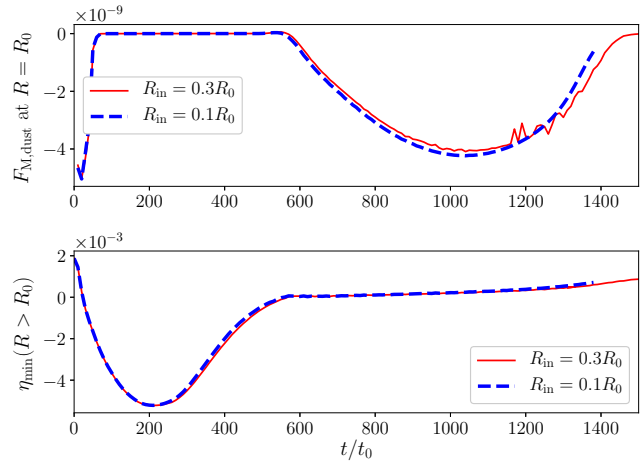}}
		\caption{
		Time variation of the dust flux at $R=R_0$ (top panel), and the minimum value of $\eta$ in the region of $R>R_0$ (bottom), in the cases with $R_{\rm in}=0.1 R_0$ and $0.3 R_0$.
		The planet mass, aspect ratio, and viscosity are the same as these in Figure~\ref{fig:epoch_vs_mineta}.
		\label{fig:dustflux_comprin}
		}
	\end{center}
\end{figure}
We set the inner boundary of the computational domain to $R_{\rm in} = 0.3 R_0$ in the fiducial case.
However, since some gaps and rings are formed at closer to the inner boundary ($\sim 0.4 R_0$), the inner boundary can affect these structures.
To check the effects, we carried out the simulation with $R_{\rm in} = 0.1 R_0$ (the location of the outer boundary, the numbers of the meshes in the radial and azimuthal directions, and other parameters are the same as those described in Section~\ref{sec:model}).
Figure~\ref{fig:dens2d_rin0.1} illustrates the two-dimensional distribution of the dust surface density, which is similar to Figure~\ref{fig:dens2d_q1e-4_h5e-2_a3e-4}.
In the case with $R_{\rm in} =0.3 R_0$ (shown in Figure~\ref{fig:dens2d_q1e-4_h5e-2_a3e-4}), the gap/ring structures on the inside of the planetary orbit vanishes at $\sim 1000 \ t_0$, but these structures remain at $t= 1200 \ t_0$ in the case with $R_{\rm in}=0.1 R_0$ shown in Figure~\ref{fig:dens2d_rin0.1}.
Hence, we found that the structures inside the planetary orbit is significantly affected by the inner boundary.
However, the structures outside the planetary orbit are not affected, and these are almost the same in Figures~\ref{fig:dens2d_q1e-4_h5e-2_a3e-4} and \ref{fig:dens2d_rin0.1}.
Figure~\ref{fig:dustflux_comprin} shows the time variations of $\fmdust$ at $R_0$ and $\etamin$ in the region of $R>R_0$, which are related to the formation and deformation of the initial dust ring.
There is no significant difference in the cases with $R_{\rm in} = 0.1 R_0$ and $0.3 R_0$.

%

\section{Ring structures induced by low-mass planets} \label{sec:rings_lowmass}
\begin{figure*}
	\begin{center}
		\resizebox{0.98\textwidth}{!}{\includegraphics{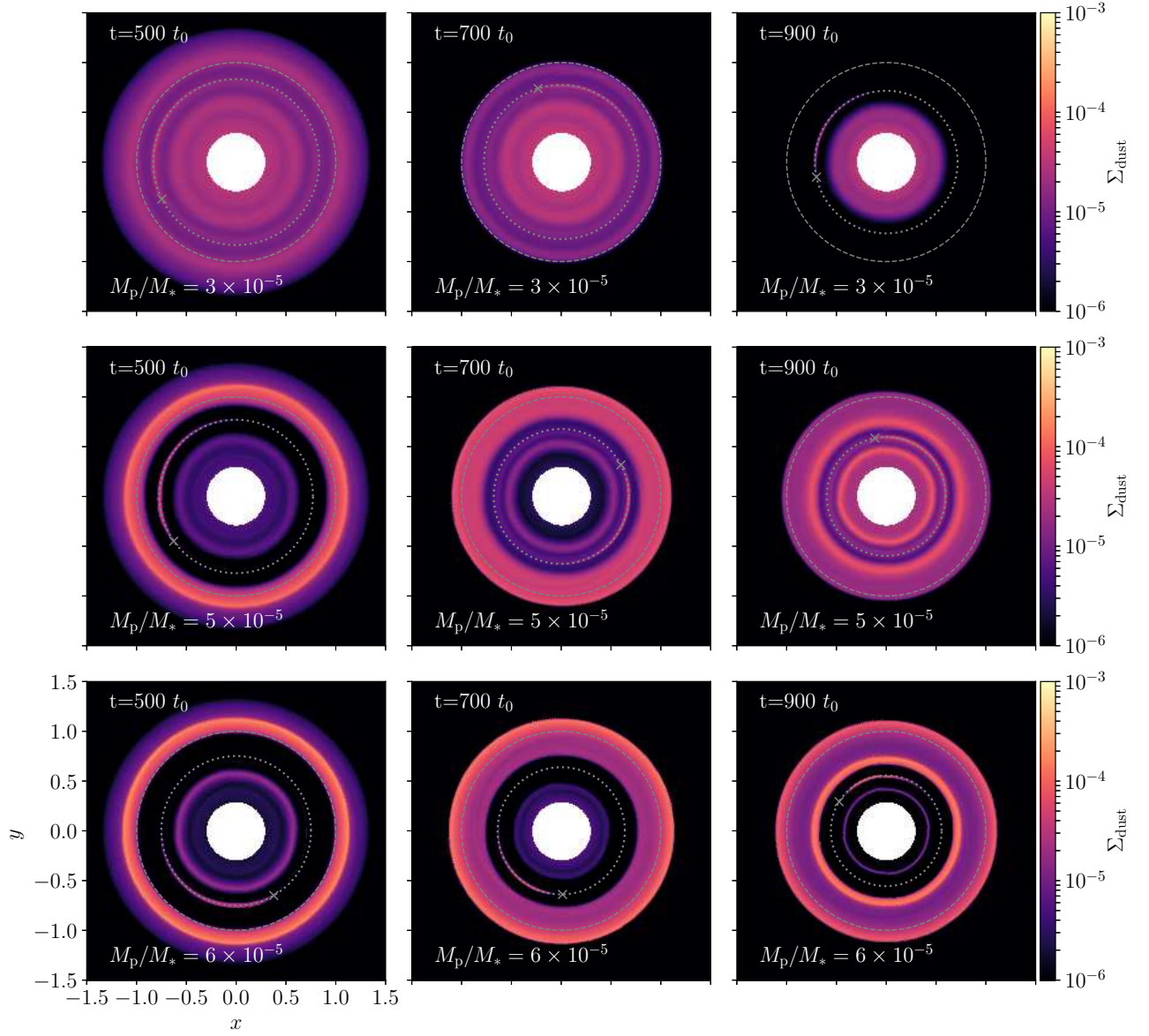}}
		\caption{
		Distributions of dust surface density in the case with $H_0=0.05$ and $\alpha=10^{-4}$.
		The masses of the planet are $\mpl/\mstar=3\times 10^{-5}$, $5\times 10^{-5}$, and $6\times 10^{-5}$, from the upper line.
		The upper case is labeled as 'no ring', and the middle and lower cases are labeled as 'follow' and 'left behind', respectively.
		The dashed and dotted circles indicate $R_0$ and $\rp$ and the cross indicates the location of the planet, as the same as these shown in Figure~\ref{fig:dens2d_rin0.1}.
		\label{fig:dens2d_lowmass}
		}
	\end{center}
\end{figure*}
We show an example of the distributions of the dust surface density in the cases labeled as 'no ring' and 'follow', in Figure~\ref{fig:dens2d_lowmass}.
In the case of $\mpl/\mstar=3\times 10^{-5}$ (\#7 in Table~\ref{tab:charatimes_mineta}, labeled as 'no ring'), since no pressure bump is formed and the dust grains drift inward faster than the planet, all the dust grains pass through the planetary orbit.
As a result, no ring structure is formed, but a shallow gap structure can form before all the dust grains go through the planetary orbit.
The cases of $\mpl/\mstar=5\times 10^{-5}$ (\#8) and $\mpl/\mstar=6\times 10^{-5}$ (\#9) are labeled as 'follow' and 'left behind', respectively.
As seen in Table~\ref{tab:charatimes_mineta}, no pressure bump is formed in the case of \#8, but the value of $\etaminovertime$ is quite close to zero, while a clear pressure bump is formed in the case of \#9.
The dust distributions in \#8 and \#9 are similar to each other, but the dust grains in \#8 are always leaking from the initial ring (because there is no pressure bump), and the ring becomes fainter quickly as compared to the case of \#9.

\begin{figure*}
	\begin{center}
		\resizebox{0.98\textwidth}{!}{\includegraphics{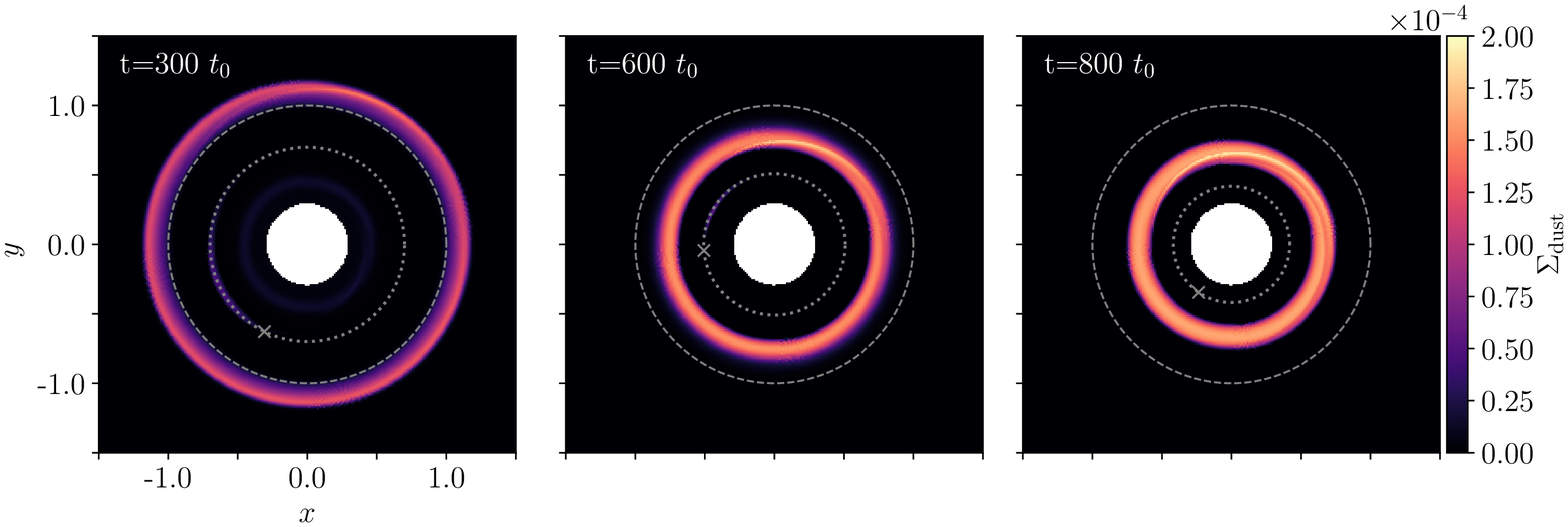}}
		\caption{
		The same as Figure~\ref{fig:dens2d_lowmass}, but for the case of $\mpl/\mstar=2\times 10^{-4}$, $H_0=0.07$, and $\alpha=3\times 10^{-4}$ (\#15 in Table~\ref{tab:charatimes_mineta}).
		This case corresponds to 'follow' case.
		\label{fig:dens2d_q2e-4_h0.07_a3e-4}
		}
	\end{center}
\end{figure*}
In Figure~\ref{fig:dens2d_q2e-4_h0.07_a3e-4}, we show the dust distributions in the case of $\mpl/\mstar=2\times 10^{-4}$, $H_0=0.07$, and $\alpha=3\times 10^{-4}$ (\#15), which is the 'follow' case.
In this case, only the last ring is visible for most of the simulation time, because the initial ring starts to be deformed at $\tstart=280 \ t_0$, and it is completely deformed at $\tend=600 \ t_0$ (see Table~\ref{tab:charatimes_mineta}).


\end{document}